\documentclass[sigconf]{acmart}

\usepackage{graphicx}
\usepackage{balance}
\usepackage{times}
\usepackage{xcolor}
\usepackage{enumitem}
\usepackage{float}
\usepackage{tikz}
\usepackage{multirow}
\usepackage{url}
\usepackage{amsmath, bbold, xspace}
\usepackage{amsthm}
\usepackage{multibib}
\usepackage{balance}
\usepackage{textcomp}
\usepackage{gensymb}
\usepackage{fancyvrb}
\usepackage{hhline}
\usepackage{setspace}
\usepackage{soul}
\usepackage{eucal}
\usepackage{etoolbox}
\usepackage{changepage}

\usepackage{eulervm}
\usepackage{algorithmicx, algorithm2e}
\usepackage{algpseudocode}

\usepackage{listings}
\definecolor{codegreen}{rgb}{0,0.6,0}
\definecolor{codegray}{rgb}{0.2,0.2,0.2}
\lstdefinestyle{pythonstyle}{
    commentstyle=\normalfont\lmr,
    commentstyle=\color{codegreen},
    keywordstyle=\textbf,
    keywordstyle=\color{blue}\textbf,
    numberstyle=\tiny\color{codegray},
    basicstyle=\scriptsize\normalfont\lmst,
    breakatwhitespace=false,
    breaklines=true,
    captionpos=b,
    keepspaces=true,
    numbersep=5pt,
    showspaces=false,
    showstringspaces=false,
    showtabs=false,
    tabsize=2,
    upquote=true,
    frame=bt,
    language=Python
}

\lstnewenvironment{python}{\lstset{style=pythonstyle}}{}
\lstdefinestyle{sqlstyle}{
    commentstyle=\color{codegreen},
    keywordstyle=\textbf,
    keywordstyle=\color{blue}\textbf,
    numberstyle=\tiny\color{codegray},
    basicstyle=\scriptsize\normalfont\lmss,
    breakatwhitespace=false,
    breaklines=true,
    captionpos=b,
    keepspaces=true,
    numbersep=5pt,
    showspaces=false,
    showstringspaces=false,
    showtabs=false,
    tabsize=2,
    upquote=true,
    frame=bt,
    language=SQL
}
\lstnewenvironment{sql}{\lstset{style=sqlstyle}}{}

\usepackage{threeparttable}

\floatname{algorithm}{Algorithm}

\makeatletter
 \def\@textbottom{\vskip \z@ \@plus 1pt}
 \let\@texttop\relax
\makeatother

\usepackage{hyphenat}
\colorlet{RED}{red}

\newcommand{\eat}[1]{}

\newcommand{\hide}[1]{} 
\newcommand{\para}[1]{\vspace{0.08in}\noindent\textbf{#1 }}

\newcommand{\lmr}{\fontfamily{lmr}\selectfont} 
\newcommand{\lmss}{\fontfamily{lmss}\selectfont} 
\newcommand{\lmst}{\fontfamily{lmtt}\selectfont} 

\newcommand{\ds}{richly formatted\xspace}

\newcommand{\systemx}{{\lmst Fonduer}\xspace}

\newcommand{\genomics}{\textsc{Genomics}\xspace}
\newcommand{\electronics}{\textsc{Electronics}\xspace}
\newcommand{\paleontology}{\textsc{Paleontology}\xspace}
\newcommand{\advertisements}{\textsc{Advertisements}\xspace}

\newcommand{\hcc}{{\lmr HasCollectorCurrent}\xspace}
\newcommand{\globalcontext}{prevalent document-level relations\xspace}
\newcommand{\GlobalContext}{Prevalent Document-Level Relations\xspace}

\newcommand{\specialcell}[2][c]{%
  \begin{tabular}[#1]{@{}c@{}}#2\end{tabular}}

\newcommand{\squishlist}{
 \begin{list}{$\bullet$}
  { \setlength{\itemsep}{0pt}
    \setlength{\parsep}{2pt}
    \setlength{\topsep}{2pt}
    \setlength{\partopsep}{0pt}
  }
}
\newcommand{\squishend}{\end{list}}

\newenvironment{nscenter}
{\parskip=8pt\par\nopagebreak\centering}
{\par\noindent\ignorespacesafterend}

\renewcommand\footnotetextcopyrightpermission[1]{} 

\usepackage[compact]{titlesec}
\titleformat{\section}{\LARGE\bfseries\uppercase}{\thesection}{1em}{}
\titleformat*{\subsection}{\Large\bfseries}
\titleformat*{\subsubsection}{\normalsize\itshape\sffamily}
\titleformat*{\paragraph}{\large\itshape}

\begin{document}

\title[Fonduer: Knowledge Base Construction from Richly Formatted
Data]{Fonduer: Knowledge Base Construction\\from Richly Formatted Data}

\author[S. Wu et al.]{
  Sen Wu\quad Luke Hsiao\quad Xiao Cheng\quad Braden Hancock\quad Theodoros Rekatsinas$^*$\\
  Philip Levis\quad Christopher R\'e
}
\affiliation{%
  \institution{Stanford University \quad $^*$University of Wisconsin-Madison \\
  \{senwu, lwhsiao, xiao, bradenjh, pal, chrismre\}@cs.stanford.edu \quad
  $^*$rekatsinas@wisc.edu}
}

\begin{abstract}
We focus on knowledge base construction (KBC) from \ds data. In contrast
to KBC from text or tabular data, KBC from \ds data aims to extract relations
conveyed jointly via textual, structural, tabular, and visual expressions. We
introduce \systemx, a machine-learning-based KBC system for \ds data. \systemx
presents a new data model that accounts for three challenging characteristics
of \ds data: (1) \globalcontext, (2) multimodality, and (3) data variety.
\systemx uses a new deep-learning model to automatically capture the
representation (i.e., features) needed to learn how to extract relations from
\ds data. Finally, \systemx provides a new programming model that enables users
to convert domain expertise, based on multiple modalities of information, to
meaningful signals of supervision for training a KBC system. \systemx-based KBC
systems are in production for a range of use cases, including at a major online
retailer. We compare \systemx against state-of-the-art KBC approaches in four
different domains. We show that \systemx achieves an average improvement of 41
F1 points on the quality of the output knowledge base---and in some cases
produces up to 1.87$\times$ the number of correct entries---compared to
expert-curated public knowledge bases. We also conduct a user study to assess
the usability of \systemx's new programming model. We show that after using
\systemx for only 30 minutes, non-domain experts are able to design KBC
systems that achieve on average $23$ F1 points higher quality than traditional
machine-learning-based KBC approaches.
\end{abstract}

%


\maketitle

\section{Introduction}
\label{sec:intro}
Knowledge base construction (KBC) is the process of populating a database with
information from data such as text, tables, images, or video. Extensive efforts
have been made to build large, high-quality knowledge bases (KBs), such as
Freebase~\cite{bollacker2008freebase}, YAGO~\cite{suchanek2008yago}, IBM
Watson~\cite{brown2013tools, ferrucci2010building},
PharmGKB~\cite{hewett2002pharmgkb}, and Google Knowledge
Graph~\cite{singhal2012introducing}. Traditionally, KBC solutions have focused
on relation extraction from unstructured text~\cite{shin2015incremental,
madaan2016numerical, nakashole2011scalable, yahya2014renoun}. These KBC systems
already support a broad range of downstream applications such as information
retrieval, question answering, medical diagnosis, and data visualization.
However, troves of information remain untapped in \emph{\ds} data, where
relations and attributes are expressed via combinations of textual, structural,
tabular, and visual cues. In these scenarios, the semantics of the data are
significantly affected by the organization and layout of the document. Examples
of \ds data include webpages, business reports, product specifications, and
scientific literature. We use the following example to demonstrate KBC from \ds
data.

\begin{figure}[t]
  \centering
  \includegraphics[width=1.0\columnwidth]{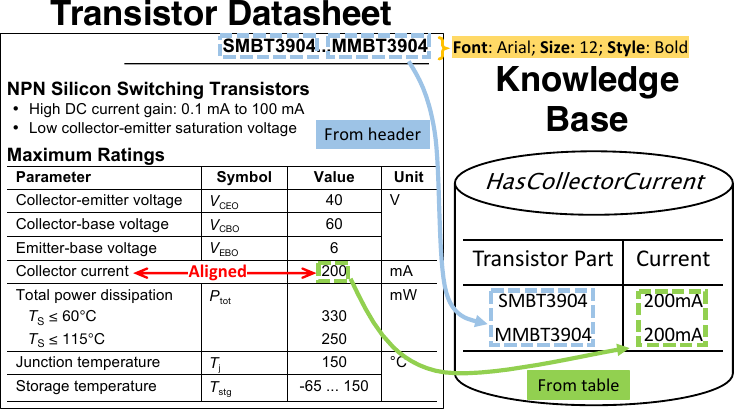}
  \caption{A KBC task to populate relation HasCollectorCurrent(Transistor Part,
    Current) from transistor datasheets. Part and Current mentions are in
    blue and green, respectively.}
  \label{fig:hardwareexp}
  \vspace{-10pt}
\end{figure}

\begin{example}[\hcc]
\label{ex:kbcex}
We highlight the \electronics domain. We are given a collection of transistor
datasheets (like the one shown in Figure~\ref{fig:hardwareexp}), and we want to
build a KB of their maximum collector currents.\footnote{Transistors are
semiconductor devices often used as switches or amplifiers. Their electrical
specifications are published by manufacturers in datasheets.} The output
KB can power a tool that verifies that transistors do not exceed
their maximum ratings in a circuit. Figure~\ref{fig:hardwareexp} shows how
relevant information is located in both the document header and table cells and
how their relationship is expressed using semantics from multiple modalities.
\end{example}

The heterogeneity of signals in \ds data poses a major challenge for existing
KBC systems. The above example shows how KBC systems that focus on text
data---and adjacent textual contexts such as sentences or paragraphs---can miss
important information due to this breadth of signals in \ds data. We review the
major challenges of KBC from \ds data.

\paragraph*{Challenges.}
KBC on \ds data poses a number of challenges beyond those present with
unstructured data:
(1) accommodating \textit{\globalcontext},
(2) capturing the \textit{multimodality} of information in the input data,
and (3) addressing the tremendous \textit{data variety}.

\para{\GlobalContext} We define the \emph{context} of a relation as the scope
information that needs to be considered when extracting the relation. Context
can range from a single sentence to a whole document. KBC systems typically
limit the context to a few sentences or a single table, assuming that relations
are expressed relatively locally. However, for \ds data, many relations rely on
information from throughout a document to be extracted.

\begin{example}[Document-Level Relations]
In Figure~\ref{fig:hardwareexp}, transistor parts are located in the document
header (boxed in blue), and the collector current value is in a table cell
(boxed in green). Moreover, the interpretation of some numerical values depends
on their units reported in another table column (e.g., 200 mA).
\label{example:intro_large_contexts}
\end{example}

\noindent
Limiting the context scope to a single sentence or table misses many potential
relations---up to 97\% in the \electronics application. On the other hand,
considering all possible entity pairs throughout the document as candidates
renders the extraction problem computationally intractable due to the
combinatorial explosion of candidates.

\para{Multimodality} Classical KBC systems model input data as
unstructured text~\cite{madaan2016numerical, shin2015incremental,
mintz2009distant}. With \ds data, semantics are part of multiple
modalities---textual, structural, tabular, and visual.

\begin{example}[Multimodality] In Figure~\ref{fig:hardwareexp}, important
information (e.g., the transistor names in the header) is expressed in larger,
bold fonts (displayed in yellow). Furthermore, the meaning of a table entry
depends on other entries with which it is visually or tabularly aligned (shown
by the red arrow). For instance, the semantics of a numeric value is specified
by an aligned unit.
\label{example:intro_multimodality}
\end{example}
\noindent Semantics from different modalities can vary significantly but can
convey complementary information.

\para{Data Variety}
With \ds data, there are two primary sources of data variety: (1) format
variety (e.g., file or table formatting) and (2) stylistic variety (e.g.,
linguistic variation).

\begin{example}[Data Variety]
In Figure~\ref{fig:hardwareexp}, numeric intervals are expressed as ``-65
\ldots 150,'' but other datasheets show intervals as ``-65 $\sim$ 150,'' or
``-65 to 150.'' Similarly, tables can be formatted with a variety of spanning
cells, header hierarchies, and layout orientations.
\end{example}

\noindent
Data variety requires KBC systems to adopt data models that are generalizable
and robust against heterogeneous input data.

\paragraph*{Our Approach.}
We introduce \systemx, a machine-learning-based system for KBC from \ds data.
\systemx takes as input \ds documents, which may be of diverse formats,
including PDF, HTML, and XML. \systemx parses the documents and analyzes the
corresponding multimodal, document-level contexts to extract relations. The
final output is a knowledge base with the relations classified to be correct.
\systemx's machine-learning-based approach must tackle a series of technical
challenges.

\para{Technical Challenges} The challenges in designing \systemx are:

\noindent (1) Reasoning about relation candidates that are manifested in
heterogeneous formats (e.g., text and tables) and span an entire document
requires \systemx's machine-learning model to analyze heterogeneous,
document-level context. While deep-learning models such as recurrent neural
networks~\cite{bahdanau2014neural} are effective with sentence- or
paragraph-level context~\cite{li2015hierarchical}, they fall short with
document-level context, such as context that span both textual and visual
features (e.g., information conveyed via fonts or
alignment)~\cite{lecun2015deep}. Developing such models is an open challenge
and active area of research~\cite{lecun2015deep}.

\noindent (2) The heterogeneity of contexts in \ds data magnifies the need for
large amounts of training data. Manual annotation is prohibitively expensive,
especially when domain expertise is required. At the same time,  human-curated
KBs, which can be used to generate training data, may exhibit low coverage or
not exist altogether. Alternatively, weak supervision sources can be used to
programmatically create large training sets, but it is often unclear how to
consistently apply these sources to \ds data. Whereas patterns in unstructured
data can be identified based on text alone, expressing patterns consistently
across different modalities in \ds data is challenging.

\noindent (3) Considering candidates across an entire document leads to a
combinatorial explosion of possible candidates, and thus random variables,
which need to be considered during learning and inference. This leads to a
fundamental tension between building a practical KBC system and learning
accurate models that exhibit high recall. In addition, the combinatorial
explosion of possible candidates results in a large class imbalance, where the
number of ``True'' candidates is much smaller than the number of ``False''
candidates. Therefore, techniques that prune candidates to balance running time
and end-to-end quality are required.

\para{Technical Contributions} Our main contributions are as follows:

\noindent(1) To account for the breadth of signals in \ds data, we design a new
{\em data model} that preserves structural and semantic information across
different data modalities. The role of \systemx's data model is twofold: (a) to
allow users to specify multimodal domain knowledge that \systemx leverages to
automate the KBC process over \ds data, and (b) to provide \systemx's
machine-learning model with the necessary representation to reason about
document-wide context (see Section~\ref{sec:overview}).

\noindent(2) We empirically show that existing deep-learning
models~\cite{zhang2016stanford} tailored for text information extraction (such
as long short-term memory (LSTM) networks~\cite{hochreiter1997long}) struggle
to capture the multimodality of \ds data. We introduce a multimodal LSTM
network that combines textual context with universal features that correspond
to structural and visual properties of the input documents. These features are
inherently captured by \systemx's data model and are generated automatically
(see Section~\ref{sec:ML_approach}). We also introduce a series of data layout
optimizations to ensure the scalability of \systemx to millions of
document-wide candidates (see Appendix~\ref{sec:trade_offs}).

\noindent(3) \systemx introduces a programming model in which no development
cycles are spent on feature engineering. Users only need to specify {\em
candidates}, the potential entries in the target KB, and provide lightweight
{\em supervision} rules which capture a user's domain knowledge and
programmatically label subsets of candidates, which are used for training
\systemx's deep-learning model (see Section~\ref{sec:multimodal_supervision}).
We conduct a user study to evaluate \systemx's programming model. We find that
when working with \ds data, users utilize the semantics from multiple
modalities of the data, including both structural and textual information in
the document. Our study demonstrates that given 30 minutes, \systemx's
programming model allows users to attain F1 scores that are $23$ points higher
than supervision via manual labeling candidates (see
Section~\ref{sec:user_study}).

\paragraph*{Summary of Results.}
\systemx-based systems are in production in a range of academic and industrial
uses cases, including a major online retailer. \systemx introduces several
advancements over prior KBC systems (see Appendix~\ref{sec:related}): (1) In
contrast to prior systems that focus on adjacent textual data, \systemx can
extract document-level relations expressed in diverse formats, ranging from
textual to tabular formats; (2) \systemx reasons about multimodal context,
i.e., both textual and visual characteristics of the input documents, to
extract more accurate relations; (3) In contrast to prior KBC systems that rely
heavily on feature engineering to achieve high quality~\cite{re2014feature},
\systemx obviates the need for feature engineering by extending a bidirectional
LSTM---the de facto deep-learning standard in natural language
processing~\cite{manning}---to obtain a representation needed to automate
relation extraction from \ds data. We evaluate \systemx in four real-world
applications of \ds information extraction and show that \systemx enables users
to build high-quality KBs, achieving an average improvement of 41 F1 points
over state-of-the-art KBC systems.

\section{Background}
\label{sec:background}
We review the concepts and terminology used in the next sections.

\subsection{Knowledge Base Construction}
The input to a KBC system is a collection of documents. The output of the
system is a relational database containing facts extracted from the input and
stored in an appropriate schema. To describe the KBC process, we adopt the
standard terminology from the KBC community. There are four types of
objects that play integral roles in KBC systems: (1) \textit{entities}, (2)
\textit{relations}, (3) \textit{mentions} of entities, and (4)
\textit{relation mentions}.

An \textbf{entity} $e$ in a knowledge base corresponds to a distinct real-world
person, place, or object. Entities can be grouped into different \textbf{entity
types} $T_1, T_2, \dots, T_n$. Entities also participate in relationships. A
relationship between $n$ entities is represented as an $n$-ary
\textbf{relation} $R(e_1, e_2, \ldots, e_n)$ and is described by a
\textbf{schema} $S_R(T_1,T_2,\ldots,T_n)$ where $e_i \in T_i$. A
\textbf{mention} $m$ is a span of text that refers to an entity. A relation
mention \textbf{candidate} (referred to as a candidate in this paper) is an
$n$-ary tuple $c=(m_1, m_2, \ldots, m_n)$ that represents a potential instance
of a relation $R(e_1, e_2, \ldots, e_n)$. A candidate classified as true is
called a \textbf{relation mention}, denoted by $r_R$.

\begin{example}[KBC]
Consider the HasCollectorCurrent task in Figure~\ref{fig:hardwareexp}. \systemx
takes a corpus of transistor datasheets as input and constructs a KB containing
the (Transistor Part, Current) binary \textbf{relation} as output. Parts like
SMBT3904 and Currents like 200mA are \textbf{entities}. The spans of text that
read ``SMBT3904'' and ``200'' (boxed in blue and green, respectively) are
\textbf{mentions} of those two entities, and together they form a
\textbf{candidate}. If the evidence in the document suggests
that these two mentions are related, then the output KB will include the
\textbf{relation mention} (SMBT3904, 200mA) of the HasCollectorCurrent
relation. \label{example:kbc} \end{example}

\noindent
The KBC problem is defined as follows:

\begin{definition}[Knowledge Base Construction]
Given a set of documents $D$ and a KB schema $S_R(T_1,T_2,\ldots,T_n)$, where
each $T_i$ corresponds to an entity type, extract a set of relation mentions
$r_R$ from $D$, which populate the schema's relational tables.
\end{definition}

\begin{figure*}[!ht]
  \centering
  \includegraphics[width=0.8\textwidth]{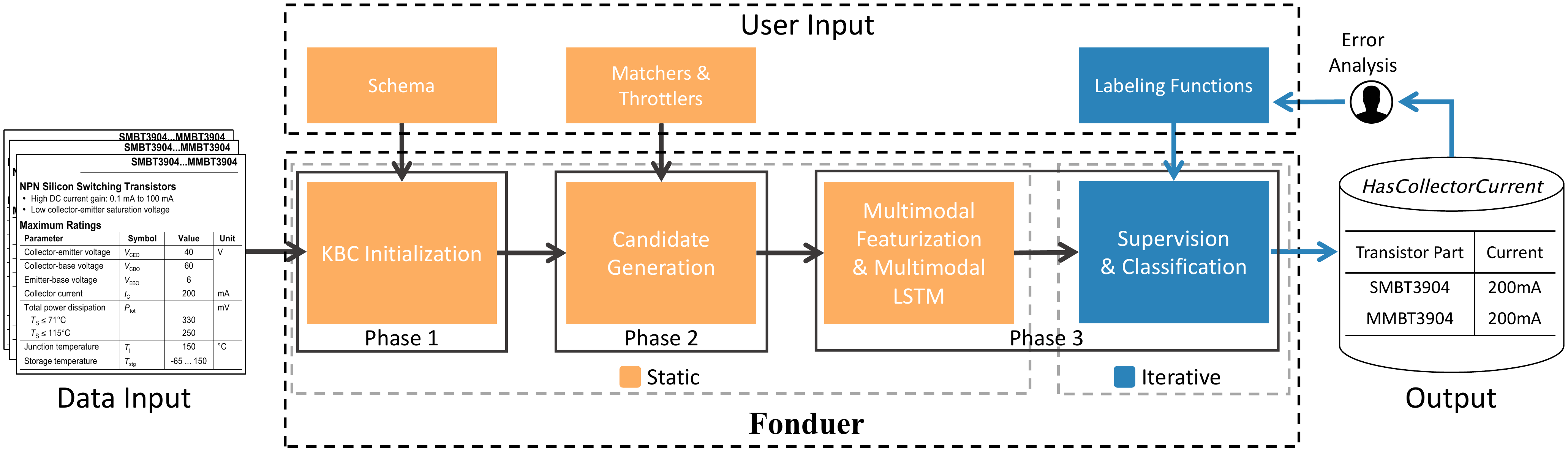}
  \caption{An overview of \systemx KBC over \ds data. Given a set of \ds
  documents and a series of lightweight inputs from the user, \systemx extracts
  facts and stores them in a relational database.}
	\label{fig:framework}
\end{figure*}

Like other machine-learning-based KBC systems~\cite{carlson2010toward,
shin2015incremental}, \systemx converts KBC to a statistical learning and
inference problem: each candidate is assigned a Boolean random variable that
can take the value ``True'' if the corresponding relation mention is correct,
or ``False'' otherwise. In machine-learning-based KBC systems, each candidate
is associated with certain features that provide evidence for the value that
the corresponding random variable should take. Machine-learning-based KBC
systems use machine learning to maximize the probability of correctly
classifying candidates, given their features and ground truth examples.

\subsection{Recurrent Neural Networks}
\label{sec:rnn_back}
The machine-learning model we use in \systemx is based on a recurrent neural
network (RNN). RNNs have obtained state-of-the-art results in many
natural-language processing (NLP) tasks, including information
extraction~\cite{graves2009novel, graves2013speech, wu2016google}. RNNs take
sequential data as input. For each element in the input sequence, the
information from previous inputs can affect the network output for the current
element. For sequential data $\{x_1, \ldots, x_T\}$, the structure of an RNN is
mathematically described as:
\begin{equation*}
  h_t = f(x_t, h_{t-1}),\qquad y = g(\{h_1, \ldots, h_T\})
\end{equation*}
where $h_t$ is the \emph{hidden state} for element $t$, and $y$ is the
\emph{representation} generated by the sequence of hidden states $\{h_1,
\ldots, h_T\}$. Functions $f$ and $g$ are nonlinear transformations. For RNNs,
we have that $f = \tanh (W_hx_t + U_hh_{t-1} + b_h)$ where $W_h$, $U_h$ are
parameter matrices and $b_h$ is a vector. Function $g$ is typically
task-specific.

\para{Long Short-term Memory} LSTM~\cite{hochreiter1997long} networks are a
special type of RNN that introduce new structures referred to as \emph{gates},
which control the flow of information and can capture long-term dependencies.
There are three types of gates: \emph{input} gates $i_t$ control which values
are updated in a memory cell; \emph{forget} gates $f_t$ control which values
remain in memory; and \emph{output} gates $o_t$ control which values in memory
are used to compute the output of the cell. The final structure of an LSTM is
given by:
\begin{align*}
  i_t &= \sigma (W_ix_t + U_ih_{t-1} + b_i) \\
  f_t &= \sigma (W_fx_t + U_fh_{t-1} + b_f) \\
  o_t &= \sigma (W_ox_t + U_oh_{t-1} + b_o) \\
  c_t &= f_t \circ c_{t-1} + i_t \circ \tanh (W_cx_t + U_ch_{t-1} + b_c)\\
  h_t &= o_t \circ \tanh (c_t)
\end{align*}
where $c_t$ is the cell state vector, $W, U, b$ are parameter matrices and a
vector, $\sigma$ is the sigmoid function, and $\circ$ is the Hadamard product.

Bidirectional LSTMs consist of forward and backward LSTMs. The forward LSTM
$f^F$ reads the sequence from $x_1$ to $x_T$ and calculates a sequence of
forward hidden states $(h_1^F, \ldots, h_T^F)$. The backward LSTM $f^B$ reads
the sequence from $x_T$ to $x_1$ and calculates a sequence of backward hidden
states $(h_1^B, \ldots, h_T^B)$. The final hidden state for the sequence is the
concatenation of the forward and backward hidden states, e.g., $h_i = [h_i^F,
h_i^B]$.

\para{Attention} Previous work explored using pooling strategies to train an
RNN, such as max pooling~\cite{verga2016multilingual}, which compresses the
information contained in potentially long input sequences to a fixed-length
internal representation by considering all parts of the input sequence
impartially. This compression of information can make it difficult for RNNs to
learn from long input sequences.

In recent years, the \emph{attention} mechanism has been introduced to overcome
this limitation by using a soft word-selection process that is conditioned on
the global information of the sentence~\cite{bahdanau2014neural}. That is,
rather than squashing all information from a source input (regardless of its
length), this mechanism allows an RNN to pay more attention to the subsets of
the input sequence where the most relevant information is concentrated.

\systemx uses a bidirectional LSTM with attention to
represent textual features of relation candidates from the documents. We
extend this LSTM with features that capture other data modalities.

\section{The {\LARGE \textbf \systemx} Framework}
\label{sec:overview}
An overview of \systemx is shown in Figure~\ref{fig:framework}. \systemx takes
as input a collection of \ds documents and a collection of user inputs. It
follows a machine-learning-based approach to extract relations from the input
documents. The relations extracted by \systemx are stored in a
target knowledge base.

We introduce \systemx's data model for representing different properties of \ds
data. We then review \systemx's data processing pipeline and describe the new
programming paradigm introduced by \systemx for KBC from \ds data.

The design of \systemx was strongly guided by interactions with collaborators
(see the user study in Section~\ref{sec:user_study}). We find that to support
KBC from \ds data, a unified data model must:

\squishlist
  \item Serve as an abstraction for system and user interaction.
  \item Capture candidates that span different areas (e.g. sections of 
    pages) and data modalities (e.g., textual and tabular data).
  \item Represent the formatting variety in richly formatted data sources in a
    unified manner.
\squishend
\systemx introduces a data model that satisfies these requirements.

\subsection{\systemx's Data Model}
\label{sec:datamodel}
\begin{figure}
  \centering
  \includegraphics[width=.8\columnwidth]{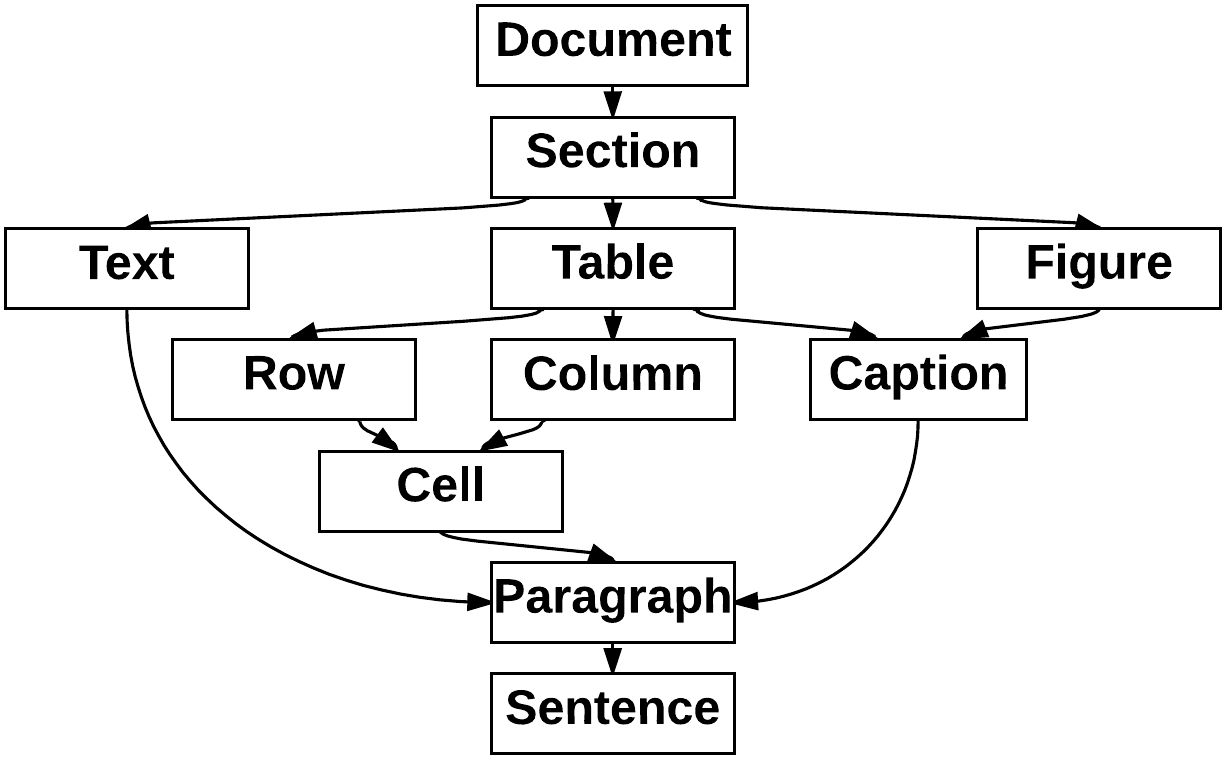}
  \vspace{-5pt}
  \caption{\systemx's data model.}
  \label{fig:data_model}
  \vspace{-17pt}
\end{figure}

\systemx's data model is a directed acyclic graph (DAG) that contains a
hierarchy of contexts, whose structure reflects the intuitive hierarchy of
document components. In this graph, each node is a \emph{context} (represented
as boxes in Figure~\ref{fig:data_model}). The root of the DAG is a
\emph{Document}, which contains \emph{Section} contexts. Each \emph{Section} is
divided into: \emph{Texts}, \emph{Tables}, and \emph{Figures}. \emph{Texts} can
contain multiple \emph{Paragraphs}; \emph{Tables} and \emph{Figures} can
contain \emph{Captions}; \emph{Tables} can also contain \emph{Rows} and
\emph{Columns}, which are in turn made up of \emph{Cells}. Each context
ultimately breaks down into \emph{Paragraphs} that are parsed into
\emph{Sentences}. In Figure~\ref{fig:data_model}, a downward edge indicates a
parent-contains-child relationship. This hierarchy serves as an abstraction for
both system and user interaction with the input corpus.

In addition, this data model allows us to capture candidates that come from
different contexts within a document. For each context, we also store the
textual contents, pointers to the parent contexts, and a wide range of
attributes from each modality found in the original document. For example,
standard NLP pre-processing tools are used to generate linguistic attributes,
such as lemmas, parts of speech tags, named entity recognition tags, dependency
paths, etc., for each \emph{Sentence}. Structural and tabular attributes of a
\emph{Sentence}, such as tags, and row/column information, and parent
attributes, can be captured by traversing its path in the data model. Visual
attributes for the document are recorded by storing bounding box and page
information for each word in a \emph{Sentence}.

\begin{example}[Data Model]
The data model representing the PDF in Figure~\ref{fig:hardwareexp} contains
one \emph{Section} with three children: a \emph{Text} for the document header,
a \emph{Text} for the description, and a \emph{Table} for the table itself
(with 10 \emph{Rows}  and 4 \emph{Columns}). Each \emph{Cell} links to both a
\emph{Row} and \emph{Column}. \emph{Text}s and \emph{Cell}s contain
\emph{Paragraphs} and \emph{Sentences}.
\end{example}

\systemx's multimodal data model unifies inputs of different formats, which
addresses the data variety of \ds data that comes from variations in format. To
construct the DAG for each document, we extract all the words in their original
order. For structural and tabular information, we use tools such as
Poppler\footnote{\url{https://poppler.freedesktop.org}} to convert an input
file into HTML format; for visual information, such as coordinates and bounding
boxes, we use a PDF printer to convert an input file into PDF format. If a
conversion occurred, we associate the multimodal information in the converted
file with all extracted words. We align the word sequences of the converted
file with their originals by checking if both their characters and number of
repeated occurrences before the current word are the same. \systemx can recover
from conversion errors by using the inherent redundancy in signals from other
modalities. In addition, this DAG structure also simplifies the variation in
format that comes from table formatting.

\paragraph*{Takeaways.}
\systemx consolidates a diverse variety of document formats, types of contexts,
and modality semantics into one model in order to address variety inherent
in \ds data. \systemx's data model serves as the formal representation of the
intermediate data utilized in all future stages of the extraction process.

\subsection{User Inputs and \systemx's Pipeline}
\label{sec:data_pipeline_overview}

The \systemx processing pipeline follows three phases. We briefly describe each
phase in turn and focus on the user inputs required by each phase. \systemx's
internals are described in Section~\ref{sec:kbc_system}.

\para{(1) KBC Initialization} The first phase in \systemx's pipeline is to
initialize the target KB where the extracted relations will be stored. During
this phase, \systemx requires the user to specify a {\em target schema}
that corresponds to the relations to be extracted. The target schema
$S_R(T_1,\ldots,T_n)$ defines a relation $R$ to be extracted from the input
documents. An example of such a schema is provided below.
\begin{example}[Relation Schema]
An example SQL schema for the relation in Figure~\ref{fig:hardwareexp} is:
\begin{sql}
CREATE TABLE HasCollectorCurrent(
  TransistorPart varchar,
  Current varchar);
\end{sql}
\end{example}

\systemx uses the user-specified schema to initialize an empty relational
database where the output KB will be stored. Furthermore, \systemx iterates
over its input {\em corpus} and transforms each document into an instance of
\systemx's data model to capture the variety and multimodality of \ds
documents.

\para{(2) Candidate Generation} In this phase, \systemx extracts relation
candidates from the input documents. Here, users are required to provide two
types of input functions: (1) \emph{matchers} and (2) \emph{throttlers}.

{\em\para{Matchers}} To generate candidates for relation $R$, \systemx requires
that users define \textit{matchers} for all distinct mention types in schema
$S_R$. Matchers are how users specify what a mention looks like. In \systemx,
matchers are Python functions that accept a span of text as input---which has
a reference to its data model---and output whether or not the match conditions
are met. Matchers range from simple regular expressions to
complicated functions that take into account signals across multiple
modalities of the input data and can also incorporate existing methods such as
named-entity recognition.

\begin{example}[Matchers]
From the HasCollectorCurrent relation in Figure~\ref{fig:hardwareexp}, users
define matchers for each type of the schema. A dictionary of valid transistor
parts can be used as the first matcher. For maximum current, users can exploit
the pattern that these values are commonly expressed as a numerical value
between 100 and 995 for their second matcher.
\begin{python}
# Use a dictionary to match transistor parts
def transistor_part_matcher(span):
  return 1 if span in part_dictionary else 0

# Use RegEx to extract numbers between [100, 995]
def max_current_matcher(span):
  return 1 if re.match('[1-9][0-9][0-5]', span) else 0
\end{python}
 \vspace{-5pt}
\end{example}

{\em \para{Throttlers}} Users can optionally provide \textit{throttlers}, which
act as hard filtering rules to reduce the number of candidates that are
materialized. Throttlers are also Python functions, but rather than accepting
spans of text as input, they operate on candidates, and output whether or not a
candidate meets the specified condition. Throttlers limit the number of
\emph{candidates} considered by \systemx.

\begin{example}[Throttler]
Continuing the example shown in Figure~\ref{fig:hardwareexp}, the user provides
a throttler, which only keeps candidates whose Current has the word ``Value''
as its column header.
\begin{python}
def value_in_column_header(cand):
  if 'Value' in header_ngrams(cand.current):
    return 1
  else:
    return 0
\end{python}
\vspace{-5pt}
\end{example}

Given the input matchers and throttlers, \systemx extracts relation candidates
by traversing its data model representation of each document. By applying
matchers to each leaf of the data model, \systemx can generate sets of mentions
for each component of the schema. The cross-product of these mentions produces
candidates:
\begin{nscenter}
	\lmss Candidate(id\textsubscript{candidate}, mention\textsubscript{1}, \ldots,
mention\textsubscript{n})
\end{nscenter}
where mentions are spans of text and contain pointers to their context in the
data model of their respective document. The output of this phase is a set of
candidates, $C$.

\para{(3) Training a Multimodal LSTM for KBC}
In this phase, \systemx trains a multimodal LSTM network to {\em classify} the
candidates generated during Phase 2 as ``True'' or ``False'' mentions of target
relations. \systemx's multimodal LSTM combines both visual and textual
features. Recent work has also proposed the use of LSTMs for KBC but has
focused only on textual data~\cite{zhang2016stanford}. In
Section~\ref{sec:feat_approach}, we experimentally demonstrate that
state-of-the-art LSTMs struggle to capture the multimodal characteristics of
\ds data, and thus, obtain poor-quality KBs.

\systemx uses a bidirectional LSTM (reviewed in Section~\ref{sec:rnn_back}) to
capture textual features and extends it with additional structural, tabular,
and visual features captured by \systemx's data model. The LSTM used by
\systemx is described in Section~\ref{sec:ML_approach}. Training in \systemx is
split into two sub-phases: (1) a multimodal featurization phase and (2) a phase
where supervision data is provided by the user.

{\em \para{Multimodal Featurization}}
Here, \systemx traverses its internal data model instance for each input
document and automatically generates features that correspond to structural,
tabular, and visual modalities as described in Section~\ref{sec:ML_approach}.
These constitute a bare-bones feature library (referred to as feature\_lib,
below), which augments the textual features learned by the LSTM. All features
are stored in a relation:

\begin{nscenter}
	\lmss Features(id\textsubscript{candidate}, LSTM\textsubscript{textual},
feature\_lib\textsubscript{others})
\end{nscenter}

\noindent No user input is required during this step. \systemx obviates the
need for feature engineering and shows that incorporating multimodal
information is key to achieving high-quality relation extraction.

{\em \para{Supervision}} To train its multimodal LSTM, \systemx requires that
users provide some form of supervision. Collecting sufficient training data for
multi-context deep-learning models is a well-established challenge. As stated
by LeCun et al.~\cite{lecun2015deep}, taking into account a context of more
than a handful of words for text-based deep-learning models requires very large
training corpora.

To soften the burden of traditional supervision, \systemx uses a supervision
paradigm referred to as {\em data programming}~\cite{ratner2016data}. Data
programming is a human-in-the-loop paradigm for training machine-learning
systems. In data programming, users only need to specify lightweight functions,
referred to as {\em labeling functions} (LFs), that programmatically assign
labels to the input candidates. A detailed overview of data programming is
provided in Appendix~\ref{sec:noise_aware_learning}. While existing work on
data programming~\cite{ratner2017snorkel} has focused on
labeling functions over textual data, \systemx paves the way for specifying
labeling functions over \ds data.

\systemx requires that users specify labeling functions that label the
candidates from Phase 2. Labeling functions in \systemx are Python functions
that take a candidate as input and assign $+1$ to label it as ``True,'' $-1$ to
label it as ``False,'' or $0$ to abstain.

\begin{example}[Labeling Functions]
\label{example:labeling_functions}
Looking at the datasheet in Figure~\ref{fig:hardwareexp}, users can express
patterns such as having the Part and Current y-aligned on the visual rendering
of the page. Similarly, users can write a rule that labels a candidate whose
Current is in the same row as the word ``current'' as ``True.''
\begin{python}{[style=pythonstyle]}
# Rule-based LF based on visual information
def y_axis_aligned(cand):
  return 1 if cand.part.y == cand.current.y else 0

# Rule-based LF based on tabular content
def has_current_in_row(cand):
  if 'current' in row_ngrams(cand.current):
    return 1
  else:
    return 0
\end{python}
\end{example}

As shown in Example~\ref{example:labeling_functions}, \systemx's internal data
model allows users to specify labeling functions that capture supervision
patterns across any modality of the data (see
Section~\ref{sec:multimodal_supervision}). In our user study, we find that it
is common for users to write labeling functions that span multiple modalities
and consider both textual and visual patterns of the input data (see
Section~\ref{sec:user_study}).

The user-specified labeling functions, together with the candidates generated
by \systemx, are passed as input to Snorkel~\cite{ratner2017snorkel}, a
data-programming engine, which converts the noisy labels generated by the input
labeling functions to denoised labeled data used to train \systemx's multimodal
LSTM model (see Appendix~\ref{sec:noise_aware_learning}).

{\em \para{Classification}}
\systemx uses its trained LSTM to assign a marginal probability to each
candidate. The last layer of \systemx's LSTM is a softmax classifier (described
in Section~\ref{sec:ML_approach}) that computes the probability of a candidate
being a ``True'' relation. In \systemx, users can specify a
\textit{threshold} over the output marginal probabilities to determine which
candidates will be classified as ``True'' (those whose marginal probability of
being true exceeds the specified threshold) and which are ``False'' (those
whose marginal probability fall beneath the threshold). This threshold depends
on the requirements of the application. Applications that require critically
high accuracy can set a high threshold value to ensure only candidates with a
high probability of being ``True'' are classified as such.

As shown in Figure~\ref{fig:framework}, supervision and classification are
typically executed over several iterations as users develop a KBC application.
This feedback loop allows users to quickly receive feedback and improve their
labeling functions, and avoids the overhead of rerunning candidate extraction
and materializing features (see Section~\ref{sec:user_study}).

\subsection{\systemx's Programming Model for KBC}
\label{sec:programming_model}

\systemx is the first system to provide the necessary abstractions and
mechanisms to enable the use of weak supervision as a means to train a KBC
system for \ds data. Traditionally, machine-learning-based KBC focuses on
feature engineering to obtain high-quality KBs. This requires that users rerun
feature extraction, learning, and inference after every modification of the
features used during KBC. With \systemx's machine-learning approach, features
are generated automatically. This puts emphasis on (1) specifying the relation
candidates and (2) providing multimodal supervision rules via labeling
functions. This approach allows users to leverage multiple sources of
supervision to address data variety introduced by variations in style better
than traditional manual labeling~\cite{shin2015incremental}.

\systemx's programming paradigm obviates the need for feature engineering and
introduces two modes of operation for \systemx applications: (1) development
and (2) production. During development, labeling functions are iteratively
improved, in terms of both coverage and accuracy, through error analysis as
shown by the blue arrows in Figure~\ref{fig:framework}. LFs are applied to a
small sample of labeled candidates and evaluated by the user on their accuracy
and coverage (the fraction of candidates receiving non-zero labels). To support
efficient error analysis, \systemx enables users to easily inspect the
resulting candidates and provides a set of labeling function metrics, such as
coverage, conflict, and overlap, which provide users with a rough assessment of
how to improve their LFs. In practice, approximately 20 iterations are adequate
for our users to generate a sufficiently tuned set of labeling functions (see
Section~\ref{sec:user_study}). In production, the finalized LFs are applied to
the entire set of candidates, and learning and inference are performed {\em
only once} to generate the final KB.

On average, only a small number of labeling functions are needed to achieve
high-quality KBC (see Section~\ref{sec:user_study}). For example, in the
\electronics application, 16 labeling functions, on average, are sufficient to
achieve an average F1 score of over 75. Furthermore, we find that tabular and
visual signals are particularly valuable forms of supervision for KBC from \ds
data, and complementary to traditional textual signals (see
Section~\ref{sec:user_study}).

\section{KBC in {\LARGE \textbf \systemx}}
\label{sec:kbc_system}
Here, we focus on the implementation of each component of \systemx. In
Appendix~\ref{sec:trade_offs} we discuss a series of optimizations that enable
\systemx's scalability to millions of candidates.

\subsection{Candidate Generation}
\label{sec:cand_gen}
Candidate generation from \ds data relies on access to document-level contexts,
which is provided by \systemx's data model. Due to the significantly
increased context needed for KBC from \ds data,  na\"{\i}vely materializing all
possible candidates is intractable as the number of candidates grows
combinatorially with the number of relation arguments. This combinatorial
explosion can lead to performance issues for KBC systems. For example, in the
\electronics domain, just 100 documents can generate over 1M candidates. In
addition, we find that the majority of these candidates do not express true
relations, creating a significant class imbalance that can hinder learning
performance~\cite{japkowicz2002class}.

To address this combinatorial explosion, \systemx allows users to specify
throttlers, in addition to matchers, to prune away excess candidates.
We find that throttlers must:
\squishlist
  \item Maintain high accuracy by only filtering negative candidates.
  \item Seek high coverage of the candidates.
\squishend
Throttlers can be viewed as a knob that allows users to trade off precision and
recall and promote scalability by reducing the number of candidates to be
classified during KBC.
\vspace{-3pt}
\begin{figure}[ht]
  \centering
  \includegraphics[width=1.\columnwidth]{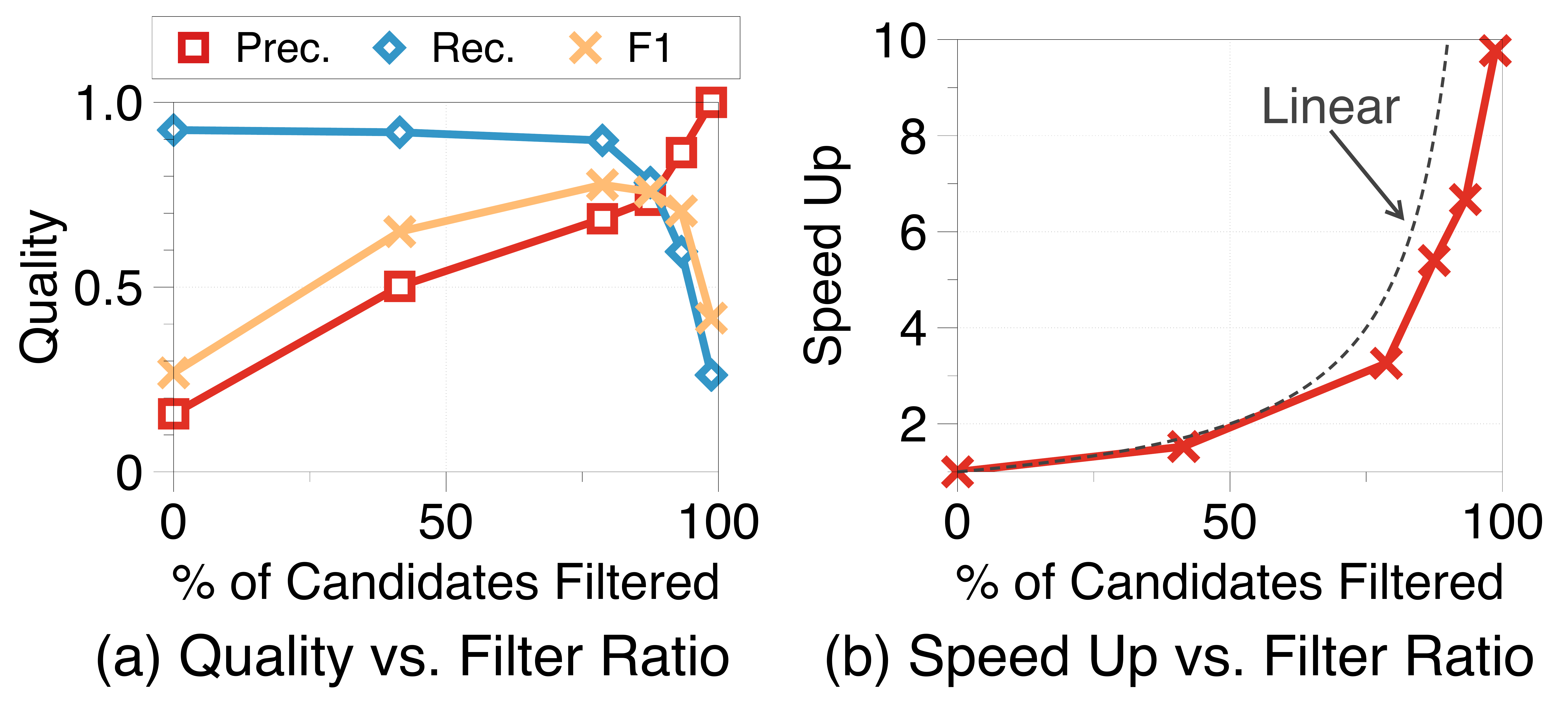}
  \caption{Tradeoff between (a) quality and (b) execution time when
  pruning the number of candidates using throttlers.}
  \label{fig:matcher_tradeoff}
\end{figure}
\vspace{-5pt}

Figure~\ref{fig:matcher_tradeoff} shows how using throttlers affects the
quality-performance tradeoff in the \electronics domain. We see that throttling
significantly improves system performance. However, increased throttling does
not monotonically improve quality since it hurts recall. This tradeoff captures
the fundamental tension between optimizing for system performance and
optimizing for end-to-end quality. When no candidates are pruned, the class
imbalance resulting from many negative candidates to the relatively small
number of positive candidates harms quality. Therefore, as a rule of thumb, we
recommend that users apply throttlers to balance negative and positive
candidates. \systemx provides users with mechanisms to evaluate this balance
over a small holdout set of labeled candidates.

\paragraph*{Takeaways.}
\systemx's data model is necessary to perform candidate generation with \ds
data. Pruning negative candidates via throttlers to balance negative and
positive candidates not only ensures the scalability of \systemx but also
improves the precision of \systemx's output.

\subsection{Multimodal LSTM Model}
\label{sec:ML_approach}

\begin{figure*}[!ht]
  \centering
  \includegraphics[width=1.75\columnwidth]{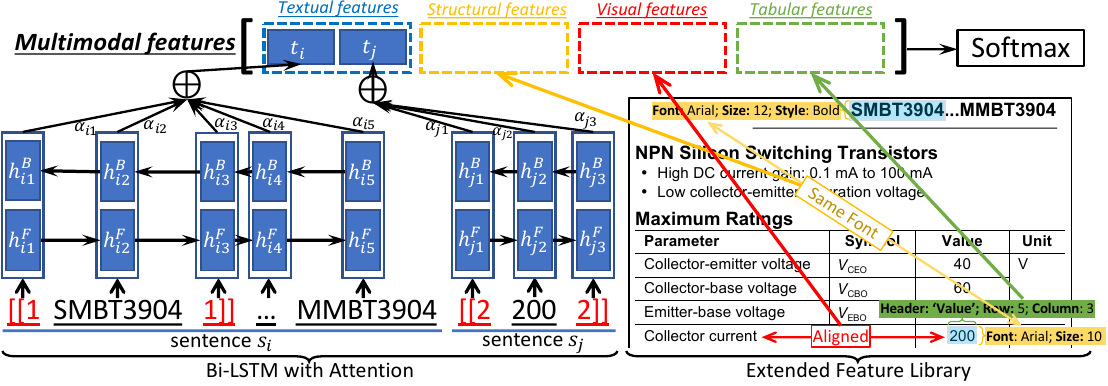}
  \caption{An illustration of \systemx's multimodal LSTM for candidate
  (SMBT3904, 200) in Figure~\ref{fig:hardwareexp}.}
  \label{fig:ML_approach_example}
    \vspace{-5pt}
\end{figure*}

We now describe \systemx's deep-learning model in detail. \systemx's model
extends a bidirectional LSTM (Bi-LSTM), the de facto deep-learning standard for
NLP~\cite{manning}, with a simple set of dynamically generated features that
capture semantics from the structural, tabular, and visual modalities of the
data model. A detailed list of these features is provided in
Appendix~\ref{sec:extended_features}. In Section~\ref{sec:quality-v-features},
we perform an ablation study demonstrating that non-textual features are key to
obtaining high-quality KBs. We find that the quality of the output KB
deteriorates up to 33 F1 points when non-textual features are removed.
Figure~\ref{fig:ML_approach_example} illustrates \systemx's LSTM. We now review
each component of \systemx's LSTM.

\para{Bidirectional LSTM with Attention} Traditionally, the primary source of
signal for relation extraction comes from unstructured text. In order to
understand textual signals, \systemx uses an LSTM network to extract
textual features. For mentions, \systemx builds a Bi-LSTM to get the
textual features of the mention from both directions of sentences
containing the candidate. For sentence $s_i$ containing the $i^{th}$ mention
in the document, the textual features $h_{ik}$ of each word $w_{ik}$
are encoded by both forward (defined as superscript $F$ in equations) and
backward (defined as superscript $B$) LSTM, which summarizes information about
the whole sentence with a focus on $w_{ik}$.
This takes the structure:
\begin{align*}
  h_{ik}^F &= LSTM(h_{i(k-1)}^F, \Phi(s_i, k)) \\
  h_{ik}^B &= LSTM(h_{i(k+1)}^B, \Phi(s_i, k)) \\
  h_{ik} &= [h_{ik}^F, h_{ik}^B]
\end{align*}
where $\Phi(s_i, k)$ is the word embedding~\cite{turian2010word}, which is
the representation of the semantics of the $k^{th}$ word in sentence $s_i$.

Then, the textual feature representation for a mention, $t_i$, is calculated by
the following attention mechanism to model the importance of different words
from the sentence $s_i$ and to aggregate the feature representation of those
words to form a final feature representation,

\begin{align*}
  u_{ik} &= \tanh(W_w h_{ik} + b_w) \\
  \alpha_{ik} &= \frac{\exp(u_{ik}^Tu_w)}{\Sigma_{j} \exp(u_{ij}^Tu_w)} \\
  t_i &= \Sigma_{j} \alpha_{ij} u_{ij}
\end{align*}

\noindent
where $W_w$, $u_w$, and $b$ are parameter matrices and a vector. $u_{ik}$ is
the hidden representation of $h_{ik}$, and $\alpha_{ik}$ is to model the
importance of each word in the sentence $s_i$. Special candidate markers (shown
in red in Figure~\ref{fig:ML_approach_example}) are added to the sentences to
draw \emph{attention} to the candidates themselves. Finally, the textual
features of a candidate are the concatenation of its mentions' textual features
$[t_1,\ldots,t_n]$.

\para{Extended Feature Library} Features for structural, tabular, and visual
modalities are generated by leveraging the data model, which preserves each
modality's semantics. For each candidate, such as the candidate (SMBT3904, 200)
shown in Figure~\ref{fig:ML_approach_example}, \systemx locates each mention in
the data model and traverses the DAG to compute features from the modality
information stored in the nodes of the graph. For example, \systemx can
traverse sibling nodes to add tabular features such as featurizing a node based
on the other mentions in the same row or column.  Similarly, \systemx can
traverse the data model to extract structural features from tags stored while
parsing the document along with the hierarchy of the document elements
themselves. We review each modality:

{\bf \textit{Structural features.}} These provide signals intrinsic to a
document's structure. These features are dynamically generated and
allow \systemx to learn from structural attributes, such as parent and sibling
relationships and XML\slash HTML tag metadata found in the
data model (shown in yellow in Figure~\ref{fig:ML_approach_example}).
The data model also allows \systemx to track structural distances of
candidates, which helps when a candidate's mentions are visually distant, but
structurally close together. Specifically, featurizing a candidate with the
distance to the lowest common ancestor in the data model is a positive signal
for linking table captions to table contents.

{\bf \textit{Tabular features.}} These are a special subset of structural
features since tables are very common structures inside documents and have high
information density. Table features are drawn from the grid-like representation
of rows and columns stored in the data model, shown in green in
Figure~\ref{fig:ML_approach_example}. In addition to the tabular location of
mentions, \systemx also featurizes candidates with signals such as
being in the same row or column. For example, consider a table that has
cells with multiple lines of text; recording that two mentions share a
row captures a signal that a visual alignment feature could easily miss.

{\bf \textit{Visual features.}} These provide signals observed from a visual
rendering of a document. In cases where tabular or structural features are
noisy---including nearly all documents converted from PDF to HTML by generic
tools---visual features can provide a complementary view of the dependencies
among text. Visual features encode many highly predictive types of semantic
information implicitly, such as position on a page, which may imply when text
is a title or header. An example of this is shown in red in
Figure~\ref{fig:ML_approach_example}.

\para{Training} All parameters of \systemx's LSTM are jointly trained,
including the parameters of the Bi-LSTM as well as the weights of
the last softmax layer that correspond to additional features.

\paragraph*{Takeaways.}
To achieve high-quality KBC with \ds data, it is vital to have features from
multiple data modalities. These features are only obtainable through traversing
and accessing modality attributes stored in the data model.

\subsection{Multimodal Supervision}
\label{sec:multimodal_supervision}
Unlike KBC from unstructured text, KBC from \ds data requires supervision from
multiple modalities of the data. In \ds data, useful patterns for KBC are more
sparse and hidden in non-textual signals, which motivates the need to exploit
overlap and repetition in a variety of patterns over multiple modalities.
\systemx's data model allows users to directly express correctness using
textual, structural, tabular, or visual characteristics, in addition to
traditional supervision sources like existing KBs. In the \electronics domain,
over 70\% of labeling functions written by our users are based on non-textual
signals. It is acceptable for these labeling functions to be noisy and conflict
with one another. Data programming theory (see Appendix~\ref{sec:dp_theory})
shows that, with a sufficient number of labeling functions, data programming
can still achieve quality comparable to using manually labeled data.

In Section~\ref{sec:lf-granularity}, we find that using metadata in the
\electronics domain, such as structural, tabular, and visual cues, results in a
66 F1 point increase over using textual supervision sources alone. Using both
sources gives a further increase of 2 F1 points over metadata alone. We also
show that supervision using information from all modalities, rather than
textual information alone, results in an increase of 43 F1 points, on average,
over a variety of domains. Using multiple supervision sources is crucial to
achieving high-quality information extraction from \ds data.

\paragraph*{Takeaways.}
Supervision using multiple modalities of \ds data is key to achieving high
end-to-end quality. Like multimodal featurization, multimodal supervision is
also enabled by \systemx's data model and addresses stylistic data variety.

\section{Experiments}
\label{sec:experiments}
We evaluate \systemx over four applications: \electronics, \advertisements,
\paleontology, and \genomics---each containing several relation extraction
tasks. We seek to answer: (1) how does \systemx compare against both
state-of-the-art KBC techniques and manually curated knowledge bases? and (2)
how does each component of \systemx contribute to end-to-end extraction
quality?

\subsection{Experimental Settings}
\label{sec:experimental}

\paragraph*{Datasets.}
The datasets used for evaluation vary in size and format.
Table~\ref{tab:dataset_sizes} shows a summary of these datasets.

\para{Electronics}
The \electronics dataset is a collection of single bipolar transistor
specification datasheets from over 20 manufacturers, downloaded from
Digi-Key.\footnote{\url{https://www.digikey.com}} These documents consist
primarily of tables and express relations containing domain-specific symbols.
We focus on the relations between transistor part numbers and several of their
electrical characteristics. \textit{We use this dataset to evaluate
\emph{\systemx} with respect to datasets that consist primarily of tables and
numerical data.}

\para{Advertisements}
The \advertisements dataset contains webpages that may contain evidence of
human trafficking activity. These webpages may provide prices of services,
locations, contact information, physical characteristics of the victims, etc.
Here, we extract all attributes associated with a trafficking advertisement.
The output is deployed in production and is used by law enforcement agencies.
This is a heterogeneous dataset containing millions of webpages over 692 web
domains in which users create customized ads, resulting in 100,000s of unique
layouts. \textit{We use this dataset to examine the robustness of
\emph{\systemx} in the presence of significant data variety.}

\para{Paleontology}
The \paleontology dataset is a collection of well-curated paleontology journal
articles on fossils and ancient organisms. Here, we extract relations between
paleontological discoveries and their corresponding physical measurements.
These papers often contain tables spanning multiple pages. Thus, achieving high
quality in this application requires linking content in tables to the text that
references it, which can be separated by 20 pages or more in the document.
\textit{We use this dataset to test \emph{\systemx}'s ability to draw
candidates from document-level contexts.}

\para{Genomics}
The \genomics dataset is a collection of open-access biomedical papers on
gene-wide association studies (GWAS) from the manually curated GWAS
Catalog~\cite{welter2014nhgri}. Here, we extract relations between
single-nucleotide polymorphisms and human phenotypes found to be statistically
significant. This dataset is published in XML format, thus, we do not have
visual representations. \textit{We use this dataset to evaluate how well the
\emph{\systemx} framework extracts relations from data that is published
natively in a tree-based format.}

\begin{table}
\footnotesize
  \begin{center}
  \caption{Summary of the datasets used in our experiments.}
  \label{tab:dataset_sizes}
    \begin{tabular}{l|rrrl}
      \hline
      \textbf{Dataset} & \textbf{Size} & \textbf{\#Docs} & \textbf{\#Rels} &
\textbf{Format} \\ \hline
      \textsc{Elec.}   &  3GB  & 7K   & 4  & PDF  \\
      \textsc{Ads.}    & 52GB  & 9.3M & 4  & HTML \\
      \textsc{Paleo.}  & 95GB  & 0.3M & 10 & PDF  \\
      \textsc{Gen.}    & 1.8GB & 589  & 4  & XML  \\ \hline
    \end{tabular}
  \end{center}
    \vspace{-15pt}
\end{table}

\paragraph*{Comparison Methods.}
We use two different methods to evaluate the quality of \systemx's output: the
upper bound of state-of-the-art KBC systems (\emph{Oracle}) and manually
curated knowledge bases (\emph{Existing Knowledge Bases}).

\para{Oracle}
Existing state-of-the-art information extraction (IE) methods focus on either
{\em textual data} or {\em semi-structured and tabular data}. We compare
\systemx against both types of IE methods. Each IE method can be split into (1)
a candidate generation stage and (2) a filtering stage, the latter of which
eliminates false positive candidates. For comparison, we approximate the upper
bound of quality of three state-of-the-art information extraction techniques by
experimentally measuring the recall achieved in the candidate generation stage
of each technique and assuming that all candidates found using a particular
technique are correct. That is, we assume the filtering stage is perfect by
assuming a precision of $1.0$.

\squishlist
  \item \textbf{Text:} We consider IE methods over
    text~\cite{shin2015incremental, madaan2016numerical}. Here, candidates are
    extracted from individual sentences, which are pre-processed with standard
    NLP tools to add part-of-speech tags, linguistic parsing information, etc.
  \item \textbf{Table:} For tables, we use an IE method for semi-structured
    data~\cite{barowy2015flashrelate}. Candidates are drawn from individual
    tables by utilizing table content and structure.
  \item \textbf{Ensemble:} We also implement an ensemble (proposed
    in~\cite{dong2014knowledge}) as the union of candidates generated by
    \textbf{Text} and \textbf{Table}.
\squishend

\para{Existing Knowledge Base}
We use existing knowledge bases as another comparison method. The \electronics
application is compared against the transistor specifications published by
Digi-Key, while \genomics is compared to both GWAS Central~\cite{beck2014gwas}
and GWAS Catalog~\cite{welter2014nhgri}, which are the most comprehensive
collections of GWAS data and widely-used public datasets. Knowledge bases such
as these are constructed using a combination of manual entry, web aggregation,
paid third-party services, and automation tools.

\paragraph*{\emph{{\large \systemx}} Details.}
\systemx is implemented in Python, with database operations being handled by
PostgreSQL. All experiments are executed in Jupyter Notebooks on a machine with
four CPUs (each CPU is a 14-core 2.40 GHz Xeon E5--4657L), 1 TB RAM, and
12$\times$3TB hard drives, with the Ubuntu 14.04 operating system.

\subsection{Experimental Results}
\label{sec:e2e-quality}

\subsubsection{Oracle Comparison}

\begin{table}[!t]
\footnotesize
  \begin{center}
  \caption{End-to-end quality in terms of precision, recall, and F1 score for
  each application compared to the upper bound of state-of-the-art
  systems.}
  \label{tab:quality}
  \vspace{-5pt}
  \begin{threeparttable}
    \begin{tabular}{l|l|rrrr}
      \hline
      \textbf{Sys.} & Metric & Text & Table & Ensemble & \systemx \\
      \hline
      \multirow{3}{*}{\textsc{Elec.}} & Prec. & 1.00 & 1.00 & 1.00 & 0.73 \\
                                      & Rec.  & 0.03 & 0.20 & 0.21 & 0.81 \\
                                      & F1    & 0.06 & 0.40 & 0.42 &
\textbf{0.77} \\
      \hline
      \multirow{3}{*}{\textsc{Ads.}} & Prec. & 1.00 & 1.00 & 1.00 & 0.87 \\
                                     & Rec.  & 0.44 & 0.37 & 0.76 & 0.89 \\
                                     & F1    & 0.61 & 0.54 & 0.86 &
\textbf{0.88} \\
      \hline
      \multirow{3}{*}{\textsc{Paleo.}} & Prec. & 0.00 & 1.00 & 1.00 & 0.72 \\
                                       & Rec.  & 0.00 & 0.04 & 0.04 & 0.38 \\
                                       & F1    & 0.00\tnote{*} & 0.08 & 0.08 &
\textbf{0.51} \\
      \hline
      \multirow{3}{*}{\textsc{Gen.}} & Prec. & 0.00 & 0.00 & 0.00 & 0.89 \\
                                     & Rec.  & 0.00 & 0.00 & 0.00 & 0.81 \\
                                     & F1    & 0.00\tnote{$^\#$} &
0.00\tnote{$^\#$} & 0.00\tnote{$^\#$} & \textbf{0.85} \\
      \hline
    \end{tabular}
    \begin{tablenotes}
      \scriptsize
      \item[*] \textit{Text} did not find any candidates.
      \item[\#] No full tuples could be created using \textit{Text} or
\textit{Table} alone
    \end{tablenotes}
  \end{threeparttable}
  \end{center}
    \vspace{-5pt}
\end{table}

We compare the end-to-end quality of \systemx to the upper bound of
state-of-the-art systems. In Table~\ref{tab:quality}, we see that \systemx
outperforms these upper bounds for each dataset. In \electronics, \systemx
results in a significant improvement of 71 F1 points over a text-only approach.
In contrast, \advertisements has a higher upper bound with text than with
tables, which reflects how advertisements rely more on text than the largely
numerical tables found in \electronics. In the \paleontology dataset, which
depends on linking references from text to tables, the unified approach of
\systemx results in an increase of 43 F1 points over the Ensemble baseline. In
\genomics, all candidates are cross-context, preventing both the text-only and
the table-only approaches from finding any valid candidates.

\subsubsection{Existing Knowledge Base Comparison}
\begin{table}[!t]
\footnotesize
  \begin{center}
  \caption{End-to-end quality vs.\ existing knowledge
bases.}\label{tab:coverage}

    \begin{tabular}{l|r|r|r}
      \hline
      \textbf{System} & \multicolumn{1}{c}{\textsc{Elec.}} &
\multicolumn{2}{|c}{\textsc{Gen.}} \\
      \hline
      Knowledge Base & Digi-Key & \specialcell{GWAS \\ Central} &
\specialcell{GWAS \\ Catalog} \\
      \hline
      \# Entries in KB & 376 & 3,008 & 4,023 \\
      \# Entries in \systemx & 447 & 6,420 & 6,420 \\
      Coverage & 0.99 & 0.82 & 0.80 \\
      Accuracy & 0.87 & 0.87 & 0.89 \\
      \# New Correct Entries & 17 & 3,154 & 2,486 \\
      Increase in Correct Entries & 1.05$\times$ & 1.87$\times$ & 1.42$\times$ \\
      \hline
    \end{tabular}
  \end{center}
    \vspace{-10pt}
\end{table}

We now compare \systemx against existing knowledge bases for \electronics and
\genomics. No manually curated KBs are available for the other two datasets. In
Table~\ref{tab:coverage}, we find that \systemx achieves high coverage of the
existing knowledge bases, while also correctly extracting novel relation
entries with over 85\% accuracy in both applications. In \electronics, \systemx
achieved 99\% coverage and extracted an additional 17 correct entries not found
in Digi-Key's catalog. In the \genomics application, we see that \systemx
provides over 80\% coverage of both existing KBs and finds 1.87$\times$ and
1.42$\times$ more correct entries than GWAS Central and GWAS Catalog,
respectively.

\paragraph*{Takeaways.}
\systemx achieves over 41 F1 points higher quality on average when compared
against the upper bound of state-of-the-art approaches. Furthermore, \systemx
attains over 80\% of the data in existing public knowledge bases while
providing up to 1.87$\times$ the number of correct entries with high accuracy.

\subsection{Ablation Studies}
We conduct ablation studies to assess the effect of context scope, multimodal
features, featurization approaches, and multimodal supervision on the quality
of \systemx. In each study, we change one component of \systemx and hold the
others constant.

\subsubsection{Context Scope Study}
\label{sec:quality-v-context}

\begin{figure}
  \centering
  \includegraphics[width=.75\columnwidth]{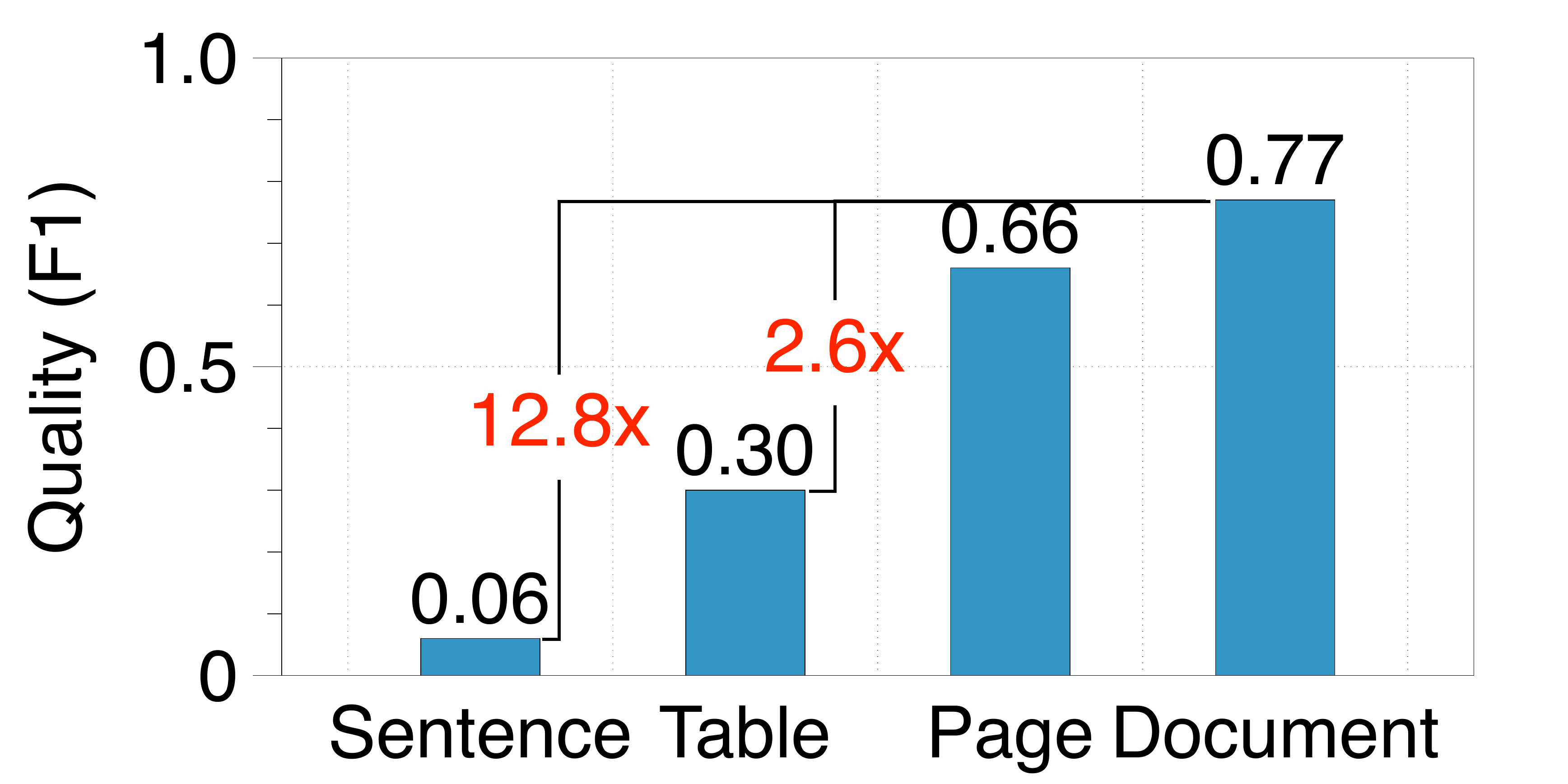}
  \vspace{-5pt}
  \caption{Average F1 score over four relations when broadening the extraction
  context scope in \electronics.}
  \label{fig:context}
  \vspace{-15pt}
\end{figure}

To evaluate the importance of addressing the non-local nature of candidates in
\ds data, we analyze how the different context scopes contribute to end-to-end
quality. We limit the extracted candidates to four levels of context scope in
\electronics and report the average F1 score for each. Figure~\ref{fig:context}
shows that increasing context scope can significantly improve the F1 score.
Considering document context gives an additional 71 F1 points (12.8$\times$)
over sentence contexts and 47 F1 points (2.6$\times$) over table contexts. The
positive correlation between quality and context scope matches our
expectations, since larger context scope is required to form candidates jointly
from both table content and surrounding text. We see a smaller increase of 11
F1 points (1.2$\times$) in quality between page and document contexts since
many of the \electronics relation mentions are presented on the first page of
the document.

\paragraph*{Takeaways.}
Semantics can be distributed in a document or implied in its structure, thus
requiring larger context scope than the traditional sentence-level contexts
used in previous KBC systems.

\subsubsection{Feature Ablation Study}
\label{sec:quality-v-features}
\begin{figure}
  \centering
  \includegraphics[width=1.\columnwidth]{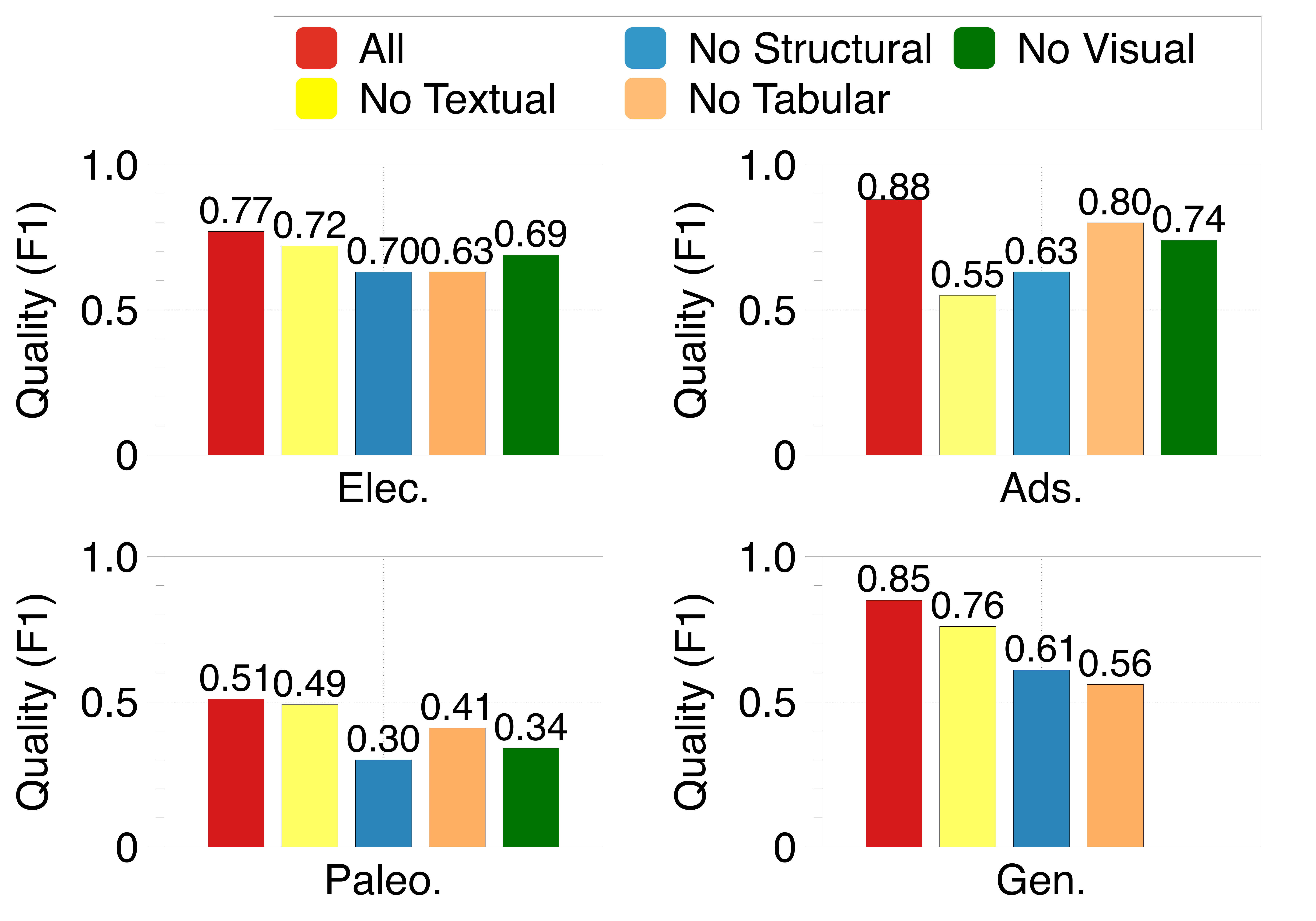}
  \vspace{-20pt}
  \caption{The impact of each modality in the feature library.}
  \label{fig:features}
    \vspace{-10pt}
\end{figure}

We evaluate \systemx's multimodal features. We analyze how different features
benefit information extraction from \ds data by comparing the effects of
disabling one feature type while leaving all other types enabled, and report
the average F1 scores of each configuration in Figure~\ref{fig:features}.

We find that removing a single feature set resulted in drops of 2 F1 points (no
textual features in \paleontology) to 33 F1 points (no textual features in
\advertisements). While it is clear in Figure~\ref{fig:features} that each
application depends on different feature types, we find that it is necessary to
incorporate all feature types to achieve the highest extraction quality.

The characteristics of each dataset affect how valuable each feature type is to
relation classification. The \advertisements dataset consists of webpages that
often use tables to format and organize information---many relations can be
found within the same cell or phrase. This heavy reliance on textual features
is reflected by the drop of 33 F1 points when textual features are disabled. In
\electronics, both components of the (part, attribute) tuples we extract are
often isolated from other text. Hence, we see a small drop of 5 F1 points when
textual features are disabled. We see a drop of 21 F1 points when structural
features are disabled in the \paleontology application due to its reliance on
structural features to link between formation names (found in text sections or
table captions) and the table itself. Finally, we see similar decreases when
disabling structural and tabular features in the \genomics application (24 and
29 F1 points, respectively). Because this dataset is published natively in XML,
structural and tabular features are almost perfectly parsed, which results in
similar impacts of these features.

\paragraph*{Takeaways.}
It is necessary to utilize multimodal features to provide a robust,
domain-agnostic description for real-world data.

\subsubsection{Featurization Study}
\label{sec:feat_approach}

\begin{table}[!t]
\footnotesize
  \begin{center}
  \caption{Comparing approaches to featurization based on \systemx's data model.}
  \label{tab:auto-manual}
  \begin{threeparttable}
    \begin{tabular}{l|l|rrr}
      \hline
      \textbf{Sys.} & Metric & Human-tuned & Bi-LSTM w/ Attn. & \systemx \\
      \hline
      \multirow{3}{*}{\textsc{Elec.}} & Prec. & 0.71 & 0.42 & 0.73\\
                                      & Rec.  & 0.82 & 0.50 & 0.81\\
                                      & F1    & 0.76 & 0.45 & \textbf{0.77} \\
      \hline
      \multirow{3}{*}{\textsc{Ads.}} & Prec. & 0.88 & 0.51 & 0.87\\
                                     & Rec.  & 0.88 & 0.43 & 0.89\\
                                     & F1    & \textbf{0.88} & 0.47 & \textbf{0.88} \\
      \hline
      \multirow{3}{*}{\textsc{Paleo.}} & Prec. & 0.92 & 0.52 & 0.76\\
                                       & Rec.  & 0.37 & 0.15 & 0.38\\
                                       & F1    & \textbf{0.53} & 0.23 & 0.51\\
      \hline
      \multirow{3}{*}{\textsc{Gen.}} & Prec. & 0.92 & 0.66 & 0.89\\
                                     & Rec.  & 0.82 & 0.41 & 0.81\\
                                     & F1    & \textbf{0.87} & 0.47 & 0.85\\
      \hline
    \end{tabular}
  \end{threeparttable}
  \end{center}
  \vspace{-10pt}
\end{table}

\begin{table}
  \footnotesize
  \centering
  \caption{Comparing the features of SRV and \systemx.}
  \label{tab:srv}
  \begin{threeparttable}
  \begin{tabular}{l|rrr}
    \hline
    \textbf{Feature Model} & Precision & Recall & F1   \\
    \hline
    SRV           & 0.72      & 0.34   & 0.39 \\
    \systemx      & 0.87      & 0.89   & 0.88 \\
    \hline
  \end{tabular}
  \end{threeparttable}
  \vspace{-5pt}
\end{table}

\begin{table}
  \footnotesize
  \centering
  \caption{Comparing document-level RNN and \systemx's deep-learning model on a
  single relation from \electronics.}
  \label{tab:learning_model}
  \begin{threeparttable}
  \begin{tabular}{l|r r}
    \hline
    \textbf{Learning Model} & Runtime during Training (secs/epoch) & Quality (F1) \\
    \hline
    Document-level RNN & 37,421 & 0.26 \\
    \systemx           & 48     & 0.65 \\
    \hline
  \end{tabular}
  \end{threeparttable}
  \vspace{-5pt}
\end{table}

We compare \systemx's multimodal featurization with: (1) a human-tuned
multimodal feature library that leverages \systemx's data model, requiring
feature engineering; (2) a Bi-LSTM with attention model; this RNN considers
textual features only; (3) a machine-learning-based system for information
extraction, referred to as SRV, which relies on HTML
features~\cite{freitag1998information}; and (4) a document-level
RNN~\cite{li2015hierarchical}, which learns a representation over all available
modes of information captured by \systemx's data model. We find that:
\squishlist
  \item \systemx's automatic multimodal featurization approach produces results
    that are comparable to manually-tuned feature representations requiring
    feature engineering. \systemx's neural network is able to extract relations
    with a quality comparable to the human-tuned approach in all datasets
    differing by no more than 2 F1 points (see Table~\ref{tab:auto-manual}).
  \item \systemx's RNN outperforms a standard, out-of-the-box Bi-LSTM
    significantly. The F1-score obtained by \systemx's multimodal RNN model is
    $1.7\times$ to $2.2\times$ higher than that of a typical Bi-LSTM  (see
    Table~\ref{tab:auto-manual}).
  \item \systemx outperforms extraction systems that leverage HTML features
    alone. Table~\ref{tab:srv} shows a comparison between \systemx and
    SRV~\cite{freitag1998information} in the \advertisements domain---the only
    one of our datasets with HTML documents as input.  \systemx's features
    capture more information than SRV's HTML-based features, which only capture
    structural and textual information. This results in $2.3\times$ higher
    quality.
  \item Using a document-level RNN to learn a single representation across all
    possible modalities results in neural networks with structures that are too
    large and too unique to batch effectively. This leads to slow runtime
    during training and poor-quality KBs. In Table~\ref{tab:learning_model}, we
    compare the performance of a document-level RNN~\cite{li2015hierarchical}
    and \systemx's approach of appending non-textual information in the last
    layer of the model. As shown \systemx's multimodal RNN obtains an F1-score
    that is almost $3\times$ higher while being three orders of magnitude
    faster to train.
\squishend
\paragraph*{Takeaways.}
Direct feature engineering is unnecessary when utilizing deep learning as a
basis to obtain the feature representation needed to extract relations from \ds
data.

\subsubsection{Supervision Ablation Study}
\label{sec:lf-granularity}

\begin{figure}
  \centering
  \includegraphics[width=.85\columnwidth]{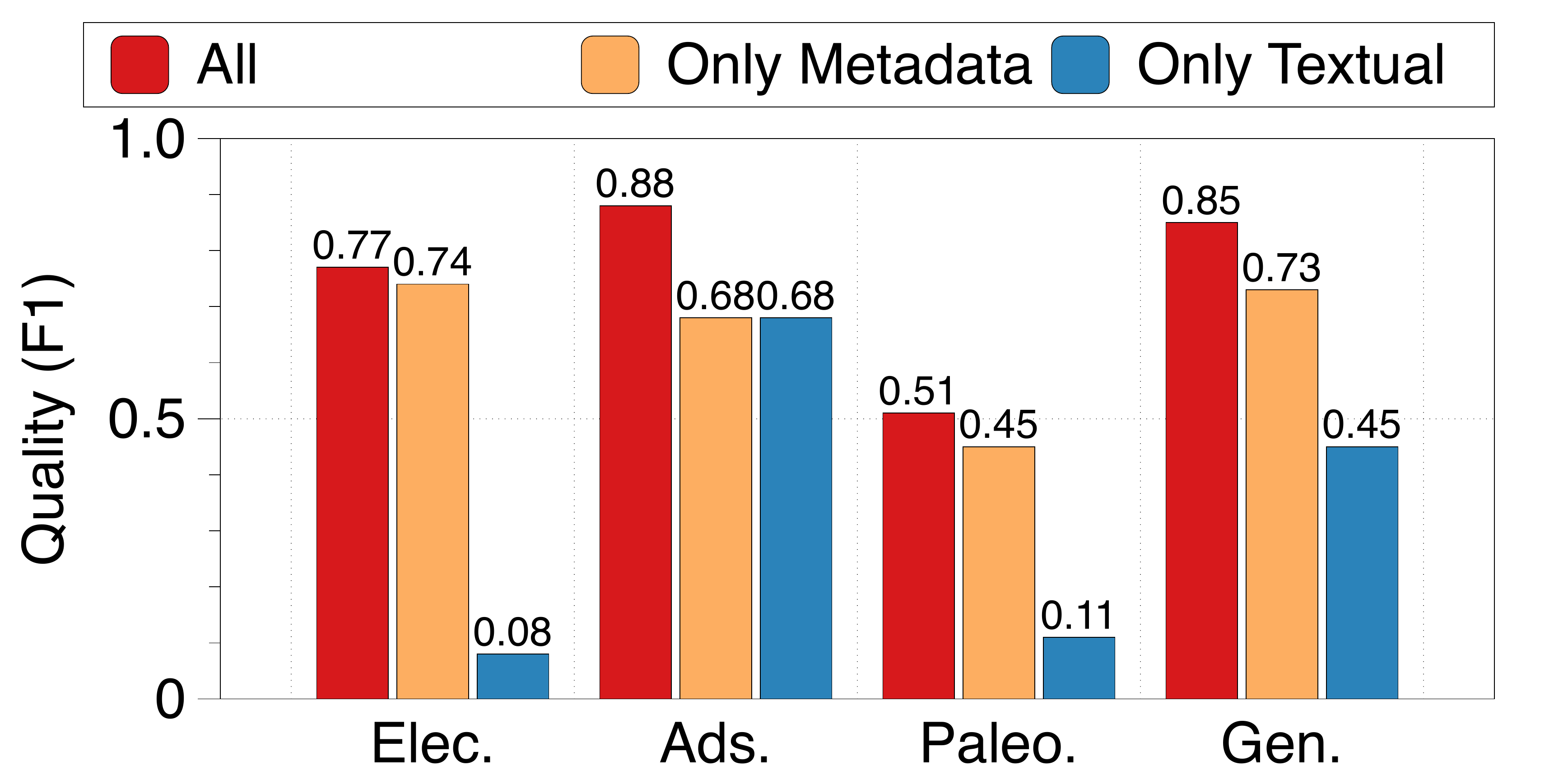}
    \vspace{-5pt}
  \caption{Study of different supervision resources on quality. Metadata
  includes structural, tabular, and visual information.}
  \label{fig:granularity}
    \vspace{-15pt}
\end{figure}

We study how quality is affected when using only textual LFs, only metadata
LFs, and the combination of the two sets of LFs. Textual LFs only operate on
textual modality characteristics while metadata LFs operate on structural,
tabular, and visual modality characteristics. Figure~\ref{fig:granularity}
shows that applying metadata-based LFs achieves higher quality than traditional
textual-level LFs alone. The highest quality is achieved when both types of LFs
are used. In \electronics, we see an increase of 66 F1 points (9.2$\times$)
when using metadata LFs and a 3 F1 point (1.04$\times$) improvement over
metadata LFs when both types are used. Because this dataset relies more heavily
on distant signals, LFs that can label correctness based on column or row
header content significantly improve extraction quality. The \advertisements
application benefits equally from metadata and textual LFs. Yet, we get an
increase of 20 F1 points (1.2$\times$) when both types of LFs are applied. The
\paleontology and \genomics applications show more moderate increases of 40
(4.6$\times$) and 40 (1.8$\times$) F1 points by using both types over only
textual LFs, respectively.

\section{User Study}
\label{sec:user_study}
Traditionally, ground truth data is created through manual annotation,
crowdsourcing, or other time-consuming methods and then used as data for
training a machine-learning model. In \systemx, we use the data-programming
model for users to programmatically generate training data, rather than needing
to perform manual annotation---a human-in-the-loop approach. In this section we
qualitatively evaluate the effectiveness of our approach compared to
traditional human labeling and observe the extent to which users leverage
non-textual semantics when labeling candidates.

We conducted a user study with 10 users, where each user was asked to complete
the relation extraction task of extracting maximum collector-emitter voltages
from the \electronics dataset. Using the same experimental settings, we compare
the effectiveness of two approaches for obtaining training data: (1) manual
annotations (Manual) and (2) using labeling functions (LF). We selected users
with a basic knowledge of Python but no expertise in the \electronics domain.
Users completed a 20 minute walk-through to familiarize themselves with the
interface and procedures. To minimize the effect of cognitive fatigue and
familiarity with the task, half of the users performed the task of manually
annotating training data first, then the task of writing labeling functions,
while the other half performed the tasks in the reverse order. We allotted 30
minutes for each task and evaluated the quality that was achieved using each
approach at several checkpoints. For manual annotations, we evaluated every
five minutes. We plotted the quality achieved by user's labeling functions each
time the user performed an iteration of supervision and classification as part
of \systemx's iterative approach. We filtered out two outliers and report
results of eight users.

\begin{figure}
  \centering
  \includegraphics[width=.85\columnwidth]{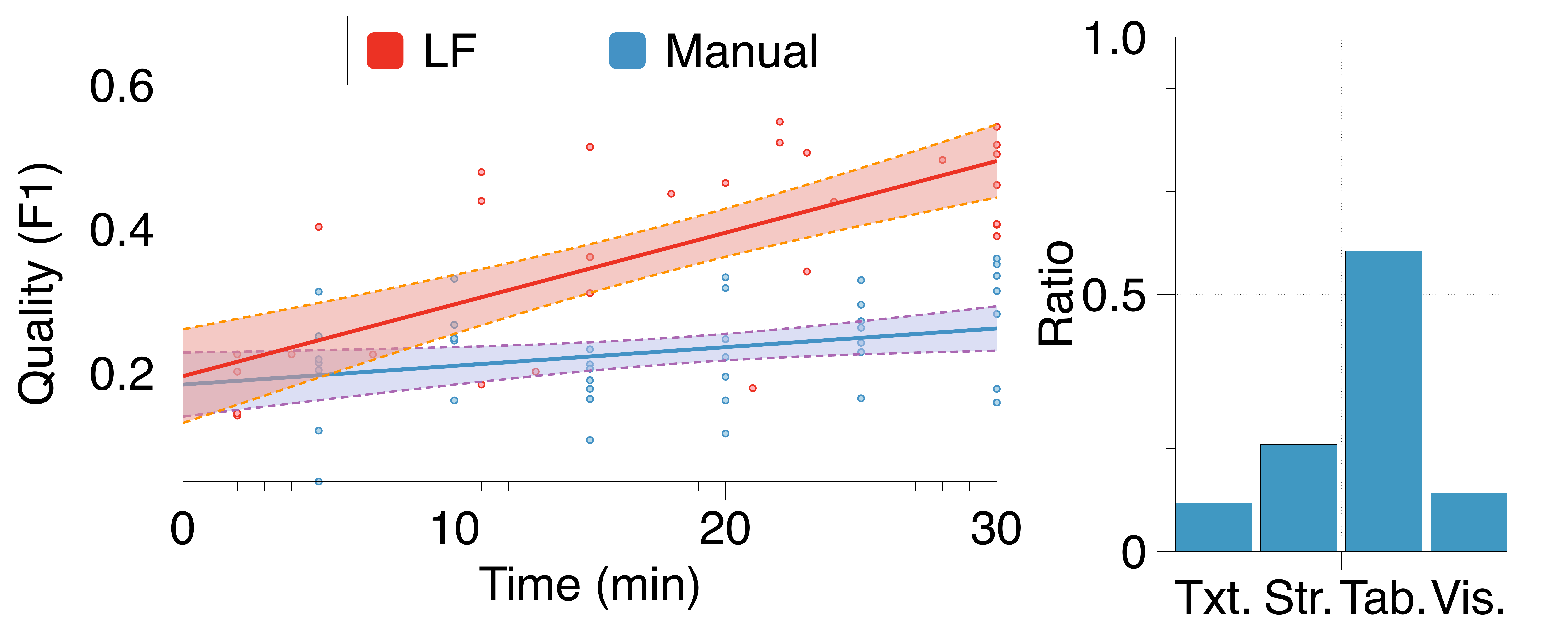}
    \vspace{-5pt}
  \caption{F1 quality over time with 95\% confidence intervals (left). Modality
    distribution of user labeling functions (right).}
  \label{fig:user_study}
    \vspace{-16pt}
\end{figure}

In Figure~\ref{fig:user_study} (left), we report the quality (F1 score)
achieved by the two different approaches. The average F1 achieved using manual
annotation was 0.26 while the average F1 score using labeling functions was
0.49, an improvement of $1.9\times$. We found with statistical significance
that all users were able to achieve higher F1 scores using labeling functions
than manually annotating candidates, regardless of the order in
which they performed the approaches. There are two primary reasons for this
trend. First, labeling functions provide a larger set of training data than
manual annotations by enabling users to apply patterns they find in the data
programmatically to all candidates---a natural desire they often vocalized
while performing manual annotation. On average, our users manually labeled 285
candidates in the allotted time, while the labeling functions they created
labeled 19,075 candidates. Users provided seven labeling functions on average.
Second, labeling functions tend to allow \systemx to learn more generic
features, whereas manual annotations may not adequately cover the
characteristics of the dataset as a whole. For example, labeling functions are
easily applied to new data.

In addition, we found that for \ds data, users relied less on textual
information---a primary signal in traditional KBC tasks---and more on
information from other modalities, as shown in Figure~\ref{fig:user_study}
(right). Users utilized the semantics from multiple modalities of the \ds data,
with 58.5\% of their labeling functions using tabular information. This
reflects the characteristics of the \electronics dataset, which contains
information that is primarily found in tables. In our study, the most common
labeling functions in each modality were:

\squishlist
\item {\bf Tabular:} labeling a candidate based on the words found in the same
  row or column.
\item {\bf Visual:} labeling a candidate based on its placement in a document
  (e.g., which page it was found on).
\item {\bf Structural:} labeling a candidate based on its tag names.
\item {\bf Textual:} labeling a candidate based on the textual characteristics
  of the voltage mention (e.g. magnitude).
\squishend

\paragraph*{Takeaways.}
We found that when working with \ds data, users relied heavily on non-textual
signals to identify candidates and weakly supervise the KBC system.
Furthermore, leveraging weak supervision allowed users to create knowledge
bases more effectively than traditional manual annotations alone.

\section{Related Work}
\label{sec:related}
We briefly review prior work in a few categories.

\para{Context Scope}
Existing KBC systems often restrict candidates to specific context scopes such
as single sentences~\cite{madaan2016numerical, yahya2014renoun} or
tables~\cite{carlson2010toward}. Others perform KBC from \ds data
by ensembling candidates discovered using separate extraction
tasks~\cite{dong2014knowledge, govindaraju2013understanding}, which overlooks
candidates composed of mentions that must be found jointly from document-level
context scopes.

\para{Multimodality}
In unstructured data information extraction systems, only textual
features~\cite{mintz2009distant} are utilized. Recognizing the need to
represent layout information as well when working with \ds data, various
additional feature libraries have been proposed. Some have relied predominantly
on structural features, usually in the context of web tables
\cite{tengli2004learning, penn2001flexible, pinto2002quasm,
freitag1998information}. Others have built systems that rely only on visual
information \cite{gatterbauer2007towards, yang2001html}. There have been
instances of visual information being used to supplement a tree-based
representation of a document \cite{kovacevic2002recognition,
cosulschi2004classifcation}, but these systems were designed for other tasks,
such as document classification and page segmentation. By utilizing our
deep-learning-based featurization approach, which supports all of these
representations, \systemx obviates the need to focus on feature engineering and
frees the user to iterate over the supervision and learning stages of the
framework.

\para{Supervision Sources}
Distant supervision is one effective way to programmatically create training
data for use in machine learning. In this paradigm, facts from existing
knowledge bases are paired with unlabeled documents to create noisy or
``weakly'' labeled training examples~\cite{mintz2009distant, min2013distant,
nguyen2011end, angeli2014stanford}. In addition to existing knowledge bases,
crowdsourcing~\cite{gao2011harnessing} and heuristics from domain experts
~\cite{pasupat2014zero} have also proven to be effective weak supervision
sources. In our work, we show that by incorporating all kinds of supervision in
one framework in a noise-aware way, we are able to achieve high quality in
knowledge base construction. Furthermore, through our programming model, we
empower users to add supervision based on intuition from any modality of data.

\section{Conclusion}
\label{sec:conclusion}

In this paper, we study how to extract information from \ds data. We show that
the key challenges of this problem are (1) \globalcontext, (2) multimodality,
and (3) data variety. To address these, we propose \systemx, the first KBC
system for \ds information extraction. We describe \systemx's data model, which
enables users to perform candidate extraction, multimodal featurization, and
multimodal supervision through a simple programming model. We evaluate \systemx
on four real-world domains and show an average improvement of 41 F1 points over
the upper bound of state-of-the-art approaches. In some domains, \systemx
extracts up to 1.87$\times$ the number of correct relations compared to
expert-curated public knowledge bases.

\begin{singlespace}
\footnotesize
\para{Acknowledgments}
We gratefully acknowledge the support of DARPA under
No. N66001-15-C-4043 (SIMPLEX),
No. FA8750-17-2-0095 (D3M),
No. FA8750-12-2-0335, and
No. FA8750-13-2-0039,
DOE 108845,
NIH U54EB020405,
ONR under No. N000141712266 and No. N000141310129,
the Intel/NSF CPS Security grant No. 1505728,
the Secure Internet of Things Project,
Qualcomm,
Ericsson,
Analog Devices,
the National Science Foundation Graduate Research Fellowship under Grant No. DGE-114747,
the Stanford Finch Family Fellowship.
the Moore Foundation,
the Okawa Research Grant,
American Family Insurance,
Accenture,
Toshiba, and
members of the Stanford DAWN project:
Intel,
Microsoft,
Google,
Teradata, and
VMware.
We thank Holly Chiang, Bryan He, and Yuhao Zhang for helpful discussions.
We also thank Prabal Dutta, Mark Horowitz, and Bj{\"o}rn Hartmann for their
feedback on early versions of this work.
The U.S. Government is authorized to reproduce and distribute reprints for
Governmental purposes notwithstanding any copyright notation thereon. Any
opinions, findings, and conclusions or recommendations expressed in this
material are those of the authors and do not necessarily reflect the views,
policies, or endorsements, either expressed or implied, of DARPA, DOE, NIH,
ONR, or the U.S. Government.
\end{singlespace}

{
\bibliographystyle{abbrv}
\bibliography{fonduer}
}

\appendix

\section{Data Programming}
\label{sec:noise_aware_learning}

Machine-learning-based KBC systems rely heavily on ground truth data (called
\textit{training data}) to achieve high quality. Traditionally, manual
annotations or incomplete KBs are used to construct training data for
machine-learning-based KBC systems. However, these resources are either costly
to obtain or may have limited coverage over the candidates considered during
the KBC process. To address this challenge, \systemx builds upon the newly
introduced paradigm of data programming~\cite{ratner2016data}, which enables
both domain experts and non-domain experts alike to programmatically generate 
large training datasets by leveraging multiple weak supervision sources and 
domain knowledge.

In data programming, which provides a framework for weak supervision, users
provide weak supervision in the form of user-defined functions, called
\emph{labeling functions}. Each labeling function provides potentially noisy
labels for a subset of the input data and are combined to create large,
potentially overlapping sets of labels which can be used to train a
machine-learning model. Many different weak-supervision approaches can be
expressed as labeling functions. This includes strategies that use existing
knowledge bases, individual annotators' labels (as in crowdsourcing), or
user-defined functions that rely on domain-specific patterns and dictionaries
to assign labels to the input data.

The aforementioned sources of supervision can have varying degrees of accuracy,
and may conflict with each other. Data programming relies on a generative
probabilistic model to estimate the accuracy of each labeling function by
reasoning about the conflicts and overlap across labeling functions. The
estimated labeling function accuracies are in turn used to assign a
probabilistic label to each candidate. These labels are used in conjunction
with a noise-aware discriminative model to train a machine-learning model for
KBC.

\subsection{Components of Data Programming}

The main components in data programming are as follows:

\para{Candidates}
A set of candidates $C$ to be probabilistically classified.

\para{Labeling Functions}
Labeling functions are used to programmatically provide labels for training
data. A labeling function is a user-defined procedure that takes a candidate as
input and outputs a label. Labels can be as simple as true or false for binary
tasks, or one of many classes for more complex multiclass tasks. Since each
labeling function is applied to all candidates and labeling functions are
rarely perfectly accurate, there may be disagreements between them. The
labeling functions provided by the user for binary classification can be more
formally defined as follows. For each labeling function $\lambda_i$ and $r \in
C$, we have $\lambda_i : r \mapsto \{-1,0,1\}$ where $+1$ or $-1$ denotes a
candidate as ``True'' or ``False'' and $0$ abstains. The output of applying a
set of $l$ labeling functions to $k$ candidates is the label matrix $\Lambda
\in \{-1,0,1\}^{k\times l}$.

\para{Output}
Data-programming frameworks output a confidence value $p$ for the
classification for each candidate as a vector $Y \in \{p\}^k$.

To perform data programming in \systemx, we rely on a data-programming engine,
Snorkel~\cite{ratner2017snorkel}. Snorkel accepts candidates and labels as
input and produces marginal probabilities for each candidate as output. These
input and output components are stored as relational tables. Their schemas are
detailed in Section~\ref{sec:overview}.

\subsection{Theoretical Guarantees}
\label{sec:dp_theory}
While data programming uses labeling functions to generate noisy training data,
it theoretically achieves a learning rate similar to methods that use manually
labeled data~\cite{ratner2016data}. In the typical supervised-learning setup,
users are required to manually label $\tilde O(\epsilon^{-2})$ examples for the
target model to achieve an expected loss of $\epsilon$. To achieve this rate,
data programming only requires the user to specify a constant number of
labeling functions that does not depend on $\epsilon$. Let $\beta$ be the
minimum coverage across labeling functions (i.e., the probability that a
labeling function provides a label for an input point) and $\gamma$ be the
minimum reliability of labeling functions, where $\gamma = 2\cdot a - 1$ with
$a$ denoting the accuracy of a labeling function. Then under the assumptions
that (1) labeling functions are conditionally independent given the true
labels of input data, (2) the number of user-provided labeling functions is at
least $\tilde O(\gamma^{-3}\beta^{-1})$, and (3) there are $k = \tilde
O(\epsilon^{-2})$ candidates, data programming achieves an expected loss
$\epsilon$. Despite the strict assumptions with respect to labeling functions,
we find that using data programming to develop KBC systems for \ds data leads
to high-quality KBs (across diverse real-world applications) even when some of
the data-programming assumptions are not met (see
Section~\ref{sec:e2e-quality}).

\section{Extended Feature Library}
\label{sec:extended_features}

\begin{table*}
  \centering
  \caption{Features from \systemx's feature library. Example values are drawn
  from the example candidate in Figure~\ref{fig:hardwareexp}. Capitalized
  prefixes represent the feature templates and the remainder of the string
  represents a feature's value.}
  \label{tab:extended_features}
  \begin{threeparttable}
    \begin{tabular}{l|l|l|l}
      \hline
      \textbf{Feature Type} & \textbf{Arity} & \textbf{Description} & \textbf{Example Value}\\
      \hline
      Structural & Unary & HTML tag of the mention & \texttt{\footnotesize TAG\_<h1>} \\
      Structural & Unary & HTML attributes of the mention & \texttt{\footnotesize HTML\_ATTR\_font-family:Arial} \\
      Structural & Unary & HTML tag of the mention's parent & \texttt{\footnotesize PARENT\_TAG\_<p>} \\
      Structural & Unary & HTML tag of the mention's previous sibling & \texttt{\footnotesize PREV\_SIB\_TAG\_<td>} \\
      Structural & Unary & HTML tag of the mention's next sibling & \texttt{\footnotesize NEXT\_SIB\_TAG\_<h1>} \\
      Structural & Unary & Position of a node among its siblings & \texttt{\footnotesize NODE\_POS\_1} \\
      Structural & Unary & HTML class sequence of the mention's ancestors & \texttt{\footnotesize ANCESTOR\_CLASS\_<s1>} \\
      Structural & Unary & HTML tag sequence of the mention's ancestors & \texttt{\footnotesize ANCESTOR\_TAG\_<body>\_<p>} \\
      Structural & Unary & HTML ID's of the mention's ancestors & \texttt{\footnotesize ANCESTOR\_ID\_l1}\tnote{b} \\
      Structural & Binary & HTML tags shared between mentions on the path to the root of the document & \texttt{\footnotesize COMMON\_ANCESTOR\_<body>} \\
      Structural & Binary & Minimum distance between two mentions to their lowest common ancestor & \texttt{\footnotesize LOWEST\_ANCESTOR\_DEPTH\_1} \\
      \hline
      Tabular & Unary & N-grams in the same cell as the mention\tnote{a} & \texttt{\footnotesize CELL\_cev}\tnote{b} \\
      Tabular & Unary & Row number of the mention & \texttt{\footnotesize ROW\_NUM\_5} \\
      Tabular & Unary & Column number of the mention & \texttt{\footnotesize COL\_NUM\_3} \\
      Tabular & Unary & Number of rows the mention spans & \texttt{\footnotesize ROW\_SPAN\_1} \\
      Tabular & Unary & Number of columns the mention spans & \texttt{\footnotesize COL\_SPAN\_1} \\
      Tabular & Unary & Row header n-grams in the table of the mention & \texttt{\footnotesize ROW\_HEAD\_collector} \\
      Tabular & Unary & Column header n-grams in the table of the mention & \texttt{\footnotesize COL\_HEAD\_value} \\
      Tabular & Unary & N-grams from all Cells that are in the same row as the given mention\tnote{a} & \texttt{\footnotesize ROW\_200\_[ma]}\tnote{c} \\
      Tabular & Unary & N-grams from all Cells that are in the same column as the given mention\tnote{a} & \texttt{\footnotesize COL\_200\_[6]}\tnote{c} \\
      Tabular & Binary & Whether two mentions are in the same table & \texttt{\footnotesize SAME\_TABLE}\tnote{b} \\
      Tabular & Binary & Row number difference if two mentions are in the same table & \texttt{\footnotesize SAME\_TABLE\_ROW\_DIFF\_1}\tnote{b} \\
      Tabular & Binary & Column number difference if two mentions are in the same table & \texttt{\footnotesize SAME\_TABLE\_COL\_DIFF\_3}\tnote{b} \\
      Tabular & Binary & Manhattan distance between two mentions in the same table & \texttt{\footnotesize SAME\_TABLE\_MANHATTAN\_DIST\_10}\tnote{b} \\
      Tabular & Binary & Whether two mentions are in the same cell & \texttt{\footnotesize SAME\_CELL}\tnote{b} \\
      Tabular & Binary & Word distance between mentions in the same cell & \texttt{\footnotesize WORD\_DIFF\_1}\tnote{b} \\
      Tabular & Binary & Character distance between  mentions in the same cell & \texttt{\footnotesize CHAR\_DIFF\_1}\tnote{b} \\
      Tabular & Binary & Whether two mentions in a cell are in the same sentence & \texttt{\footnotesize SAME\_PHRASE}\tnote{b} \\
      Tabular & Binary & Whether two mention are in the different tables & \texttt{\footnotesize DIFF\_TABLE}\tnote{b} \\
      Tabular & Binary & Row number difference if two mentions are in different tables & \texttt{\footnotesize DIFF\_TABLE\_ROW\_DIFF\_4}\tnote{b} \\
      Tabular & Binary & Column number difference if two mentions are in different tables & \texttt{\footnotesize  DIFF\_TABLE\_COL\_DIFF\_2}\tnote{b} \\
      Tabular & Binary & Manhattan distance between two mentions in different tables & \texttt{\footnotesize DIFF\_TABLE\_MANHATTAN\_DIST\_7}\tnote{b} \\
      \hline
      Visual & Unary & N-grams of all lemmas visually aligned with the mention\tnote{a} & \texttt{\footnotesize ALIGNED\_current} \\
      Visual & Unary & Page number of the mention & \texttt{\footnotesize PAGE\_1} \\
      Visual & Binary & Whether two mentions are on the same page & \texttt{\footnotesize SAME\_PAGE} \\
      Visual & Binary & Whether two mentions are horizontally aligned & \texttt{\footnotesize HORZ\_ALIGNED}\tnote{b} \\
      Visual & Binary & Whether two mentions are vertically aligned & \texttt{\footnotesize VERT\_ALIGNED} \\
      Visual & Binary & Whether two mentions' left bounding-box borders are vertically aligned & \texttt{\footnotesize  VERT\_ALIGNED\_LEFT}\tnote{b} \\
      Visual & Binary & Whether two mentions' right bounding-box borders are vertically aligned & \texttt{\footnotesize VERT\_ALIGNED\_RIGHT}\tnote{b} \\
      Visual & Binary & Whether the centers of two mentions' bounding boxes are vertically aligned & \texttt{\footnotesize VERT\_ALIGNED\_CENTER}\tnote{b} \\
      \hline
      \hline
    \end{tabular}
    \begin{tablenotes}
      \item[a] All N-grams are 1-grams by default.
      \item[b] This feature was not present in the example candidate. The
        values shown are example values from other documents.
      \item[c] In this example, the mention is 200, which forms part of the
        feature prefix. The value is shown in square brackets.
    \end{tablenotes}
  \end{threeparttable}

\end{table*}

\systemx augments a bidirectional LSTM with features from an extended feature
library in order to better model the multiple modalities of \ds data. In
addition, these extended features can provide signals drawn from large
contexts since they can be calculated using \systemx's data model of the
document rather than being limited to a single sentence or table. In
Section~\ref{sec:experiments}, we find that including multimodal features is
critical to achieving high-quality relation extraction. The provided extended
feature library serves as a baseline example of these types of features that
can be easily enhanced in the future. However, even with these baseline
features, our users have been able to build high-quality knowledge bases for
their applications.

The extended feature library consists of a baseline set of features from the
structural, tabular, and visual modalities. Table~\ref{tab:extended_features}
lists the details of the extended feature library.
Features are represented as strings, and each feature space is then mapped into
a one-dimensional bit vector for each candidate, where each bit represents
whether the candidate has the corresponding feature.

\section{{\LARGE \textbf \systemx} at Scale}
\label{sec:trade_offs}

We use two optimizations to enable \systemx's scalability to millions of
candidates (see Section~\ref{sec:datamodel}): (1) data caching and (2) data
representations that optimize data access during the KBC process. Such
optimizations are standard in database systems. Nonetheless, their impact on
KBC has not been studied in detail.

Each candidate to be classified by \systemx's LSTM as ``True'' or ``False'' is
associated with a set of mentions (see
Section~\ref{sec:data_pipeline_overview}). For each candidate, \systemx's
multimodal featurization generates features that describe each individual
mention in isolation and features that jointly describe the set of all mentions
in the candidate. Since each mention can be associated with many different
candidates, we cache the featurization of each mention. Caching during
featurization results in a $100\times$ speed-up on average in the \electronics
domain yet only accounts for $10\%$ of this stage's memory usage.

Recall from Section~\ref{sec:programming_model} that \systemx's programming
model introduces two modes of operation: (1) development, where users
iteratively improve the quality of labeling functions without executing the
entire pipeline; and (2) production, where the full pipeline is executed once
to produce the knowledge base. We use different data representations to
implement the abstract data structures of {\lmss Features} and {\lmss Labels}
(a structure that stores the output of labeling functions after applying them
over the generated candidates). Implementing {\lmss Features} as a
list-of-lists structure minimizes runtime in both modes of operation since it
accounts for sparsity. We also find that {\lmss Labels} implemented as a
coordinate list during the development mode are optimal for fast updates. A
list-of-lists implementation is used for {\lmss Labels} in production mode.

\subsection{Data Caching}
With \ds data, which frequently requires document-level context, thousands of
candidates need to be featurized for each document. Candidate features from the
extended feature library are computed at both the mention level and relation
level by traversing the data model accessing modality attributes. Because
each mention is part of many candidates, na\"{\i}ve featurization
of candidates can result in the redundant computation of thousands of mention
features. This pattern highlights the value of data caching when performing
multimodal featurization on \ds data.

Traditional KBC systems that operate on single sentences of unstructured text
pragmatically assume that only a small number of candidates will need to be
featurized for each sentence and do not cache mention features as a result.

\begin{example}[Inefficient Featurization]
\label{example:inefficient_featurization}
In Figure~\ref{fig:hardwareexp}, the transistor part mention MMBT3904 could be
matched with up to 15 different numerical values in the datasheet. Without
caching, the features of the MMBT3904 would be unnecessarily recalculated 14
times, once for each candidate. In real documents 100s of feature calculations
would be wasted.
\end{example}

\noindent
In Example~\ref{example:inefficient_featurization}, eliminating unnecessary
feature computations can improve performance by an order of magnitude.

To optimize the feature-generation process, \systemx implements a
document-level caching scheme for mention features. The first computation of a
mention feature requires traversing the data model. Then, the result is cached
for fast access if the feature is needed again. All features are cached until
all candidates in a document are fully featurized, after which the cache is
flushed. Because \systemx operates on documents atomically, caching a single
document at a time improves performance without adding significant memory
overhead. In the \electronics application, we find that caching achieves over
100$\times$ speed-up on average and in some cases even over 1000$\times$, while
only accounting for approximately 10\% of the memory footprint of
the featurization stage.

\paragraph*{Takeaways.}
When performing feature generation from \ds data, caching the intermediate
results can yield over 1000$\times$ improvements in featurization runtime
without adding significant memory overhead.

\subsection{Data Representations}
\label{sec:tradeoff}

The \systemx programming model involves two modes of operation: (1) development
and (2) production. In development, users iteratively improve the quality of
their labeling functions through error analysis and without executing the full
pipeline as in previous techniques such as incremental
KBC~\cite{shin2015incremental}. Once the labeling functions are finalized, the
\systemx pipeline is only run once in production.

In both modes of operation, \systemx produces two abstract data structures
({\lmss Features} and {\lmss Labels} as described in
Section~\ref{sec:overview}). These data structures have three access patterns:
(1) \emph{materialization}, where the data structure is created; (2)
\emph{updates}, which include inserts, deletions, and value changes; and (3)
\emph{queries}, where users can inspect the features and labels to make
informed updates to labeling functions.

Both {\lmss Features} and {\lmss Labels} can be viewed as matrices, where each
row represents annotations for a candidate (see
Section~\ref{sec:data_pipeline_overview}). {\lmss Features} are dynamically
named during multimodal featurization but are static for the lifetime of a
candidate. {\lmss Labels} are statically named in classification but
updated during development. Typically {\lmss Features} are sparse: in the
\electronics application, each candidate has about 100 features while the
number of unique features can be more than 10M. {\lmss Labels} are also
sparse, where the number of unique labels is the number of labeling
functions.

The data representation that is implemented to store these abstract data
structures can significantly affect overall system runtime. In the \electronics
application, multimodal featurization accounts for 50\% of end-to-end runtime,
while classification accounts for 15\%. We discuss two common sparse matrix
representations that can be materialized in a SQL database.

\squishlist
  \item \textbf{List of lists} (LIL): Each row stores a list of (column\_key,
    value) pairs. Zero-valued pairs are omitted. An entire row can be retrieved
    in a single query. However, updating values requires iterating over
    sublists.
  \item \textbf{Coordinate list} (COO): Rows store (row\_key, column\_key,
    value) triples. Zero-valued triples are omitted. With COO, multiple queries
    must be performed to fetch a row's attributes. However, updating values
    takes constant time.
\squishend

The choice of data representation for {\lmss Features} and {\lmss Labels}
reflects their different access patterns, as well as the mode of operation.
During development, {\lmss Features} are materialized once, but frequently
queried during the iterative KBC process. {\lmss Labels} are updated each time
a user modifies labeling functions. In production, {\lmss Features}' access
pattern remains the same. However, {\lmss Labels} are not updated once users
have finalized their set of labeling functions.

From the access patterns in the \systemx pipeline, and the characteristics of
each sparse matrix representation, we find that implementing {\lmss Features}
as an LIL minimizes runtime in production and development. {\lmss Labels},
however, should be implemented as COO to support fast insertions during
iterative KBC and reduce runtimes for each iteration. In production, {\lmss
Labels} can also be implemented as LIL to avoid the computation overhead of
COO. In the \electronics application, we find that LIL provides $1.4\times$
speed-up over COO in production and that COO provides over $5.8\times$
speed-up over LIL when adding a new labeling function.

\paragraph*{Takeaways.}
We find that {\lmss Labels} should be implemented as a coordinate list during
development, which supports fast updates for supervision, while {\lmss
Features} should use a list of lists, which provides faster query times. In
production, both {\lmss Features} and {\lmss Labels} should use a list-of-list
representation.

\section{Future Work}
\label{sec:future}
Being able to extract information from \ds data enables a wide range of
applications, and represents a new and interesting research direction. While we
have demonstrated that \systemx can already obtain high-quality knowledge bases
in several applications, we recognize that many interesting challenges remain.
We briefly discuss some of these challenges.

\para{Data Model} One challenge in extracting information from \ds data comes
directly at the data level---we cannot perfectly preserve all document
information. Future work in parsing, OCR, and computer vision have the
potential to improve the quality of \systemx's data model for complex table
structures and figures. For example, improving the granularity of
\systemx's data model to be able to identify axis titles, legends, and
footnotes could provide additional signals to learn from and additional
specificity for users to leverage while using the \systemx programming model.

\para{Deep-Learning Model} \systemx's multimodal recurrent neural network
provides a prototypical automated featurization approach that achieves high
quality across several domains. However, future developments for incorporating
domain-specific features could strengthen these models. In addition,
it may be possible to expand our deep-learning model to perform additional
tasks (e.g., identifying candidates) to simplify the \systemx pipeline.

\para{Programming Model} \systemx currently exposes a Python interface to allow
users to provide weak supervision. However, further research in user interfaces
for weak supervision could bolster user efficiency in \systemx. For example,
allowing users to use natural language or graphical interfaces in supervision
may result in improved efficiency and reduced development time
through a more powerful programming model. Similarly, feedback techniques
like \emph{active learning}~\cite{settles1648active} could empower users to
more quickly recognize classes of candidates that need further disambiguation
with LFs.

\end{document}